\documentclass[%
 reprint,
 superscriptaddress,
nofootinbib,
 amsmath,amssymb,
 aps,
prd,
]{revtex4-2}

\usepackage{graphicx}
\usepackage{dcolumn}
\usepackage{bm}
\usepackage{xcolor}
\usepackage{orcidlink}

\usepackage{cleveref}

\crefname{figure}{Fig.}{Figs.}
\Crefname{figure}{Figure}{Figures}
\crefname{table}{Tab.}{Tabs.}
\Crefname{table}{Table}{Tables}
\crefname{section}{Sec.}{Secs.}
\Crefname{section}{Section}{Sections}
\crefname{equation}{Eq.}{Eqs.}
\Crefname{equation}{Equation}{Equations}
\crefname{appendix}{App.}{Apps.}
\Crefname{appendix}{Appendix}{Appendices}

\newcommand{\Lfree}{\Delta L_\mathrm{free}}
\newcommand{\Lres}{\Delta L_\mathrm{res}}
\newcommand{\Lctrl}{\Delta L_\mathrm{ctrl}}
\newcommand{\Rmodel}{R^\mathrm{(model)}}
\newcommand{\Msun}{M_\odot}
\newcommand{\Mdet}{M_\mathrm{f}}
\newcommand{\MIMR}{M_\mathrm{f}^\mathrm{IMR}}

\begin{document}

\preprint{APS/123-QED}

\title{Accounting for detector calibration uncertainty in black-hole ringdown analysis}

\author{Rowina S. Nathan\,\orcidlink{0000-0002-3922-2773}}
\email{rowina.nathan@anu.edu.au}
\affiliation{OzGrav-ANU, Centre for Gravitational Astrophysics,
Research School of Physics and Research School of Astronomy \& Astrophysics, The Australian National University, ACT 2601, Australia}
\author{Shingo Hido\,\orcidlink{0009-0009-0004-4170}}%
\email{shingo@icrr.u-tokyo.ac.jp}
\affiliation{KAGRA Observatory, Institute for Cosmic Ray Research, The University of
Tokyo, 238 Higashi-Mozumi, Kamioka-cho, Hida City, Gifu 506-1205, Japan}%
\author{Neil Lu\,\orcidlink{0000-0002-8861-9902}}
\affiliation{OzGrav-ANU, Centre for Gravitational Astrophysics,
Research School of Physics and Research School of Astronomy \& Astrophysics, The Australian National University, ACT 2601, Australia}
\author{Mallika R. Sinha\,\orcidlink{0009-0008-0906-6328}}
\affiliation{OzGrav-Monash, School of Physics and Astronomy, Monash University, Clayton VIC 3800, Australia}
\author{Sizheng Ma\,\orcidlink{0000-0002-4645-453X}}
\affiliation{Perimeter Institute for Theoretical Physics, Waterloo, ON N2L2Y5, Canada}
\author{Ling Sun\,\orcidlink{0000-0001-7959-892X}}
\affiliation{OzGrav-ANU, Centre for Gravitational Astrophysics,
Research School of Physics and Research School of Astronomy \& Astrophysics, The Australian National University, ACT 2601, Australia}

\date{\today}

\begin{abstract}
Detector calibration systematics can bias measurements of a remnant black hole's mass and spin, producing spurious evidence against general relativity if unaccounted for.
The ringdown phase in the gravitational-wave signal produced by a binary black-hole merger encodes the remnant properties: at sufficiently late times, the signal is well described by a superposition of quasinormal modes, each with a frequency and damping time set only by the remnant mass and spin. Here we present two complementary approaches for analyzing the ringdown portion of the data to estimate the remnant mass and spin in the presence of a nonnegligible detector calibration error, using the quasinormal-mode rational filter (QNMRF). With a well-measured error, we apply the best-estimated calibration correction to the strain before conducting the standard analysis; with an unknown error, we instead jointly sample the remnant mass, spin and the frequency-dependent calibration error, marginalizing over an uncertainty prior.
By injecting the numerical-relativity waveform SXS:BBH:0305 into colored detector noise, at a full-signal matched-filter signal-to-noise ratio of $\sim60$ and with simulated calibration errors of $\pm25\%$ in magnitude and $\pm25^\circ$ in phase, marginalizing over the calibration uncertainty recovers the true mass and spin within the 90\% credible interval, whereas the estimated values are biased if the calibration errors are ignored. 
We apply the method to GW250207, an event recorded while the Hanford detector response was not well characterized, and recover a remnant consistent with the inspiral-merger-ringdown prediction using the ringdown portion alone, both by correcting for a best-estimated calibration error and by marginalizing over a broad uncertainty prior.
\end{abstract}


\maketitle

\section{\label{sec:inroduction}Introduction}

Since the first detection in 2015 \cite{150914}, we have observed hundreds of gravitational-wave (GW) events \cite{GWTC5}, allowing us to infer the properties of astrophysical black holes with unprecedented precision \cite{Abac_2026_TGR, Abac_2025_250114, Abac2026_230814, Abac2025_231123, Abac2025_241011_241110}.
These events are detected using an international network of ground-based interferometric GW detectors~\cite{ObservingScenarios}: LIGO \cite{aLIGO}, Virgo \cite{aVIRGO} and KAGRA \cite{KAGRA}. 
This large catalog of events has been used to perform rigorous tests of general relativity (GR) in the strong-field regime \cite{TGR_GWTC3, TGR_GWTC4_1, TGR_GWTC4_2, TGR_GWTC4_3, TGR_GWTC5}. The signal-to-noise ratio (SNR) is a limiting factor in these tests, but as detectors advance and SNR improves~\cite{Capote2025, Soni2024}, systematics other than noise (such as waveform-modeling errors, microlensing and calibration errors \cite{Gupta_2024, Kopty_2026}) will become limiting in tests of GR. 

GW detectors do not directly measure strain; the detector response is modeled via a process known as \emph{calibration} \cite{Calib_aligo-2017, Cahillane_2017, Sun2020_calib, Sun2021_calib, Calib_virgo_2022, Calib_kagra_2021, Wade2025, Chen_2025, Virgo_calib_2026}; the raw detector output is converted into GW strain data.  This response is not always perfectly modeled: systematic errors and statistical uncertainties in the calibration model propagate into the strain data and, in turn, into the astrophysical inferences drawn from it. 
The effects of calibration errors on many phenomena have been studied, including search-pipeline sensitivity \cite{Lindblom_2009, Milan_2026}, compact-binary-coalescence (CBC) parameter estimation \cite{Payne_2020, Vitale_2012, Essick_2022, Wade2023}, population inference and cosmological measurements \cite{Huang2025}, and tests of GR \cite{Hall_2019, Gupta_2024, Sinha2025}. Under current calibration error and uncertainty estimates, most of these analyses find the impact small compared to astrophysical statistical uncertainty \cite{Milan_2026, Vitale_2012, Huang2025}, though it can become significant for sufficiently loud signals or observations with next-generation detectors \cite{Essick_2022,Evans_2021_CE}, and may affect tests of GR if left uncorrected \cite{Gupta_2024, Sinha2025}.

Following a binary black-hole (BBH) merger, the remnant black hole emits GWs as it settles into its final stationary state, a process known as ringdown \cite{Berti_2026_review}. GR places tight constraints on the GWs emitted during ringdown: at late times the signal is described by a linear superposition of quasinormal modes (QNMs) \cite{Vishveshwara_1970, Teukolsky_1973, Chandrasekhar_1975, Leaver_1985, Kokkotas_1999, Berti_2009} whose frequencies and damping times are entirely determined by the mass and spin of the remnant black hole \cite{Israel_1967, Carter1971, Hawking1972, Robinson1975}. 
Measuring more than one mode therefore gives multiple, independent estimates of the remnant mass and spin, so their mutual consistency provides a test of the Kerr hypothesis and of GR itself, an approach known as black-hole spectroscopy \cite{Dreyer_2004, Berti_2006}. The first overtone of GW150914 has been tested in this way \cite{Isi_2019, Ma2023,Lu2025_rf}, and multi-mode ringdown tests now form part of standard analyses by the LIGO--Virgo--KAGRA (LVK) Collaboration \cite{TGR_GWTC5,Abac_2025_250114,Abac_2026_TGR}.

Many analysis methods exist to measure QNMs in data including direct fitting in the time domain \cite{Carullo_2019, Isi_2019, Isi_2021, Gennari_2024, Siegel_2025_ringdownII, Morosaki_2025, Khusid_2026, Suzuki_2026}, wavelet-based analyses performed in the frequency domain \cite{Finch_2021}, gating-and-inpainting approaches \cite{Wang_2025}, parameterized extensions to effective-one-body (EOB) waveform models \cite{Brito_2018, Ghish_2021, Maggio_2023, Toubiana_2024, Pompili_2025, Grimaldi_2026}, and machine-learning-based techniques \cite{Pacilio_2024, Iida_2026}. The QNM rational filter (QNMRF) \cite{Ma_2022, Ma_2023QNM, Ma2023, Lu2025_rf} takes an amplitude-independent approach: rather than fitting a sum of damped sinusoids to the data, it constructs a filter in the frequency domain that removes a specified QNM. Because each mode's frequency and damping time are fixed by the remnant mass and spin under the no-hair theorem, the filter depends only on these two parameters, and multi-mode filters can be applied to remove additional modes or overtones without needing their amplitudes or phases known in advance.

Like most of these ringdown-only methods (which use only the post-merger segment of the signal), QNMRF has to date assumed a fixed, accurate calibration.\footnote{The exception is the parameterized-EOB approach \cite{Maggio_2023, Pompili_2025, Grimaldi_2026}, which extends a full inspiral-merger-ringdown waveform model and in principle inherits that model's calibration marginalization.} Accounting for calibration uncertainty is routine in full inspiral-merger-ringdown (IMR) parameter estimation (PE), where the frequency-dependent amplitude and phase errors are marginalized over using a spline model constrained by the detector calibration uncertainty envelopes \cite{Farr_2014_calib, GWTC4_spline, Abbott_2016_spline}. 
Astrophysical signals have long been proposed as an independent calibration reference \cite{Schutz_1986,Essick_Holz_2019, GWTC4}, and have recently been used to obtain the first informative calibration measurement \cite{astro_calib_paper}. This gap is the motivation for the present work.

Here we present two approaches for accounting for calibration errors in ringdown analysis, building on the QNMRF framework: (1) applying a well-estimated calibration correction directly within QNMRF, and (2) when the calibration error is poorly constrained, jointly sampling the remnant mass, spin, and calibration parameters to infer the remnant properties and calibration errors simultaneously. This is the first implementation of such an approach within a ringdown-only analysis framework. We validate the pipeline on numerical relativity (NR) injections and apply it to GW250207\_115645 (GW250207), an event recorded while the LIGO Hanford detector was not yet fully calibrated \cite{astro_calib_paper}. The remainder of this paper is organized as follows: \cref{sec:calib_unc} describes the calibration model, \cref{sec:filter_MCMC} the QNMRF and joint likelihood, \cref{sec:sims} the simulation validation, \cref{sec:results} the application to GW250207 and finally \cref{sec:conclusion} concluding remarks.

\section{\label{sec:calib_unc}Calibration Uncertainty}

Ground-based GW detectors are interferometers that infer spacetime perturbations by measuring minute changes in distance along their orthogonal arms. Reconstructing the detector's dimensionless strain output from a modeled detector response is a process known as calibration \cite{Calib_aligo-2017, Cahillane_2017, Sun2020_calib, Sun2021_calib, Calib_virgo_2022, Calib_kagra_2021, Wade2025, Chen_2025, Virgo_calib_2026}. The dimensionless detector strain data $d$ is defined as $d\equiv \Lfree/L$, where $\Lfree$ is the uncontrolled, free differential arm length change, and $L$ is the total arm length. However, $\Lfree$ is not measured directly, as the change in the differential arm length is suppressed by the feedback control loop. The residual displacement in differential arm length, $\Lres$, is an analog measurement that is converted into a digital error signal with a sensing transfer function $C$ such that\footnote{Here, we denote all quantities in the frequency domain.}
\begin{equation}
    \label{eq:derr}
    d_\mathrm{err}=C\,\Lres.
\end{equation}
Digital filters $D$ are then applied to $d_\mathrm{err}$ to produce the digital control signal
\begin{equation}
    \label{eq:dctrl}
    d_\mathrm{ctrl}=D\,d_\mathrm{err}.
\end{equation}
This is then fed back through the actuation path to produce an analog control displacement $\Lctrl$ to suppress the movement of the arms through the actuation transfer function $A$:
\begin{equation}
    \label{eq:Lctrl}
    \Lctrl=A\,d_\mathrm{ctrl}.
\end{equation}
With $\Lfree=\Lctrl+\Lres$, combining \crefrange{eq:derr}{eq:Lctrl} gives the detector response function $R$
\begin{equation}
    \label{eq:detector_response}
    \Lfree=\left[\frac{1}{C}+AD\right]d_\mathrm{err}=R\,d_\mathrm{err}.
\end{equation}

The true response $R$ is not known exactly, so a modeled response $\Rmodel$ is used when reconstructing the strain data. The residual calibration error is the time-dependent, frequency-dependent ratio $\eta(f;t) \equiv R(f;t)/\Rmodel(f;t)$, which departs from unity in both amplitude and phase. State-of-the-art calibration in the fourth observing run (O4)~\cite{Virgo_O4} achieves uncertainties of roughly $10\%$ in amplitude and $10^\circ$ in phase across the full sensitive band \cite{Wade2025, Chen_2025, LVK_CalUnc_O4, Virgo_calib_2026}.

Calibration parameters vary on timescales of hours to days set by the drift of the instrument \cite{Viets_2018, Sun2020_calib}, far longer than the duration of the data segment used in either a full CBC analysis or a ringdown-only analysis, typically seconds and a few milliseconds respectively, and longer than their separation from the off-source data used to estimate the noise. Calibration error is therefore effectively static across a single-event analysis, and we suppress the explicit $t$-dependence for the remainder of this work, writing $\eta(f) \equiv \eta(f;t)$.

Calibration errors are modeled and corrected using this complex correction factor \cite{Allen1996_calib, Sun2020_calib, astro_calib_paper, Sinha2025}
\begin{equation}
    \label{eq:calib}
    \eta(f)=\frac{R(f)}{R^{(\rm{model})}(f)}=\left[1+\delta\mathcal{A}(f)\right]\exp\left[i\delta\phi(f)\right],
\end{equation}
where $\delta\mathcal{A}(f)$ is the amplitude error and $\delta\phi(f)$ is the phase error. In practice, $\delta\mathcal{A}(f)$ and $\delta\phi(f)$ are estimated by combining full-band measurements taken out of observing mode with real-time monitoring at selected frequencies during observing~\cite{Sun2020_calib,Wade2025}. They are estimated and reported, at each frequency, as a median value together with a $\pm1\sigma$ uncertainty envelope on an hourly cadence~\cite{Viets_2018, LVK_CalUnc_O4}, with each frequency bin treated as independently Gaussian distributed about the median. This correction factor can then be applied to the strain data to obtain the corrected strain
\begin{equation}
    \label{eq:calib_strain}
    d^{\rm{(corr)}}=\eta d.
\end{equation}
While applying an estimated correction factor is standard practice in full inspiral-merger-ringdown parameter estimation \cite{Farr_2014_calib, GWTC4_spline, Abbott_2016_spline}, doing so within the ringdown-only QNMRF pipeline is new to this work (\cref{ssec:calib_corr}).

Since the response $R$ characterizes how the instrument converts its raw output into recorded strain, it acts identically on every contribution to the detector output, irrespective of origin: both the astrophysical signal and the detector noise pass through the same sensing and actuation chain described by Eqs.~\eqref{eq:derr}--\eqref{eq:detector_response}. Any error in the modeled response therefore modulates both by the same common factor: the recorded data $d$ contain $(h+n)/\eta$ rather than $h+n$, where $h$ and $n$ denote the signal and additive noise, respectively. Thus, the residual calibration error that distorts the signal also distorts the noise, a property we exploit when constructing the likelihood in \cref{ssec:joint_analysis}.

The level of calibration error required to bias the recovered remnant properties depends on its location in frequency relative to the ringdown modes: errors well separated from a mode's frequency leave the recovered mass and spin largely unaffected, while comparable errors near the mode frequency can substantially bias the inferred mass and spin \cite{Sinha2025}. This is a particular concern for next-generation detectors such as Cosmic Explorer \cite{Evans_2021_CE} and Einstein Telescope \cite{Maggiore_2020_ET}: Ref.~\cite{Sinha2025} finds that an uncorrected calibration error of $10\%$ can bias the recovered mass and spin by more than $3\sigma$,
excluding the true values, and that errors must be kept below roughly $4\%$ to avoid such a bias.

\section{\label{sec:filter_MCMC}Calibration correction and marginalization}

\subsection{\label{ssec:qnmrf}Ringdown analysis with quasinormal mode rational filters}

QNMs are characterized by their angular components $(\ell,m)$ and an overtone component $n$ such that their strain can be written as 
\begin{equation}
    \label{eq:qnm}
    h(t) = A_{\ell m n }e^{-(t-t_0)/\tau_{\ell m n }}\cos\left[2\pi f_{\ell m n }(t-t_0)+\phi_{\ell m n }\right],
\end{equation}
where $t_0$ is a reference time, $f_{\ell m n }$ is the frequency of oscillation for the mode, and  $\tau_{\ell m n }$ is the damping time. 

The QNMRF method works by removing QNMs in the frequency domain. The filter~\cite{Ma_2022,Ma2023,Ma_2023QNM} 
\begin{equation}
    \label{eq:filter}
    \mathcal{F}_{\ell m n }(\omega;M_\mathrm{f},\chi_\mathrm{f})=\frac{\omega-\omega_{\ell m n }}{\omega-\omega^*_{\ell m n }}\frac{\omega+\omega^*_{\ell m n }}{\omega+\omega_{\ell m n }},
\end{equation}
removes the $(\ell, m, n )$ QNM from the data where $\omega$ is the real-valued frequency, $\omega_{\ell m n }=2\pi f_{\ell m n }-i/\tau_{\ell m n }$ is the complex-valued QNM frequency. Note that this filter does not depend on the amplitude $A_{\ell m n }$ or phase $\phi_{\ell m n }$ of the corresponding individual modes, allowing for sampling over remnant mass\footnote{Masses are quoted in the detector frame throughout this work.} $M_\mathrm{f}$ and spin $\chi_\mathrm{f}$ only. Conveniently, a number of QNMs can be removed at once using a total filter,
\begin{equation}
    \label{eq:filtertot}
    \mathcal{F}_\mathrm{tot}(\omega;M_\mathrm{f},\chi_\mathrm{f})=\prod_{\ell m n }\mathcal{F}_{\ell m n }(\omega;M_\mathrm{f},\chi_\mathrm{f}).
\end{equation}

In order to apply this QNMRF to the data, the time-domain strain data must be transformed to the frequency domain. This is done via a fast Fourier transform (FFT) such that
\begin{equation}
    \label{eq:FFT}
    \tilde{d}_f=\int d_t e^{2\pi i f t}dt,
\end{equation}
where $d_t$ is the time-domain data and $\tilde{d}_f$ is the frequency domain data. We can then apply the total filter from \cref{eq:filtertot} and recover the time-domain filtered data
\begin{equation}
    \label{eq:FFT_filter}
    d_t^F=\int df \tilde{d}_f\mathcal{F}_{\rm{tot}}(f)e^{-i2\pi ft}.
\end{equation}

To estimate the remnant mass $M_\mathrm{f}$ and spin $\chi_\mathrm{f}$, we construct rational filters that remove the ringdown signal, leaving filtered data consistent with pure noise when the correct remnant parameter values are used. The likelihood can be determined with
\begin{equation}
    \label{eq:likelihood}
    \ln\mathcal{L}(d_t|M_\mathrm{f},\chi_\mathrm{f},t_0,\mathcal{F}_{\rm{tot}})=-\frac{1}{2}\sum_{i,j>I_0}d^F_iC^{-1}_{ij}d^F_j,
\end{equation}
where $i,j$ index discrete time samples, $d^F_i$ are the filtered data, $I_0$ is the sample index corresponding to the selected ringdown start time $t_0$, so that the sum runs only over data after $t_0$, and $C_{ij}$ is the noise autocovariance matrix \cite{Ma2023}.
Typically the QNMRF is applied over a grid of masses and spins, to find the mass and spin that maximizes the likelihood ($\ln \mathcal{L}$). This is referred to throughout this paper as the Grid method or QNMRF Grid. 

\subsection{Applying calibration correction}
\label{ssec:calib_corr}
Calibration estimates are not always applied to the data before parameter estimation. Where an estimate of the error is available, typically from detector measurements~\cite{Sun2020_calib, Wade2025} or occasionally inferred astrophysically in loud events~\cite{astro_calib_paper}, it can be applied to correct the strain data before the standard QNMRF analysis of \cref{ssec:qnmrf}.

Because the rational filter already acts in the frequency domain, the calibration correction enters as a multiplicative factor on the frequency-domain data in the same step as the filter, so that \cref{eq:FFT_filter} becomes
\begin{equation}
    \label{eq:filtered_corr}
    d_t^F=\int df\ \tilde{d}_f\,\eta(f)\,\mathcal{F}_{\rm{tot}}(f)\,e^{-i2\pi ft},
\end{equation}
with the frequency-dependent $\eta(f)$ fixed at its maximum-a-posteriori (MAP) estimate. The Grid method of \cref{ssec:qnmrf} is then applied to the corrected data using the standard likelihood of \cref{eq:likelihood}, unchanged.

As discussed in \cref{sec:calib_unc}, the detector response acts on all contributions in the strain, including both the signal and noise. Therefore, the same correction is applied to the pure noise segment used to compute the autocovariance matrix $C_{ij}$.

Because $\eta$ is held fixed, its uncertainty is not propagated into the recovered mass and spin. Propagating that uncertainty instead requires sampling the calibration jointly with the remnant parameters (see \cref{ssec:joint_analysis}).

\subsection{\label{ssec:joint_analysis}Full marginalization likelihood}

Equation~\eqref{eq:likelihood}, when maximized, recovers the remnant mass and spin under the
assumption that the detector calibration is known to be accurate. To account for calibration uncertainty, we model the calibration error with a set of parameters and sample it jointly with $M_\mathrm{f}$ and $\chi_\mathrm{f}$. We describe the frequency-dependent error $\eta(f)$ of \cref{eq:calib} by a small number ($K$) of control points $\{f_k\}_{k=1}^{K}$, with amplitude and phase offsets
$\{\delta\mathcal{A}_k,\,\delta\phi_k\}$ interpolated across frequency to give the
continuous profiles $\delta\mathcal{A}(f)$ and $\delta\phi(f)$. The full
parameter vector is
\begin{equation}
    \label{eq:theta}
    \bm{\theta} = \big(M_
    \mathrm{f},\ \chi_\mathrm{f},\ \{\delta\mathcal{A}_k\}_{k=1}^{K},\ 
    \{\delta\phi_k\}_{k=1}^{K}\big).
\end{equation}
This control-point parametrization with spline interpolation mirrors the spline-based calibration marginalization already used in full IMR parameter estimation \cite{Farr_2014_calib, GWTC4_spline, Abbott_2016_spline, astro_calib_paper}, here restricted to the narrow frequency band spanned by the ringdown. We parameterize the frequency-dependent calibration error with seven control points per detector; the first and last are pinned to zero, leaving $K=5$ free amplitude and $K=5$ free phase points that are sampled. These are then interpolated using a quadratic spline. The total dimension is therefore $n_{\rm dim} = 2 + 2K = 12$. 

Since the calibration error $\eta(f)$ affects both the signal and noise, there are two consequences for the likelihood. First, the parameterized calibration correction $\eta(f;\bm{\theta})$ multiplies the data in the frequency domain, so the filtered data of \cref{eq:FFT_filter} become
\begin{equation}
    \label{eq:filtered_cal}
    d^F_t(\bm{\theta}) = \int df\ \tilde{d}_f\,
    \mathcal{F}_{\rm tot}(f;M_\mathrm{f},\chi_\mathrm{f})\,\eta(f;\bm{\theta})\,e^{-i2\pi f t}.
\end{equation}
Second, because the same factor affects the noise, the noise power spectrum inherits a dependence on the amplitude error,
\begin{equation}
    \label{eq:psd_scaling}
    S(f) \longrightarrow |\eta(f;\bm{\theta})|^2\,S(f)
    = \big[1+\delta\mathcal{A}(f)\big]^2 S(f),
\end{equation}
and the noise autocovariance matrix is recomputed accordingly,
$C_{ij}\to C_{ij}(\{\delta\mathcal{A}_k\})$. Note that the covariance depends only on the amplitude error, since $|\eta|$ is independent of the phase $\delta\phi$. This rescaling is not optional: if only the data were corrected while $S(f)$ were held fixed, shrinking $|\eta|$ would shrink the residual and inflate the apparent likelihood without improving the true $M_\mathrm{f},\chi_\mathrm{f}$ fit. Conversely, because we multiply both the filtered data and the noise by the same factor at every frequency, an overall rescaling of $\eta$ (for any amplitude-error shape, not only a uniform shift) leaves the likelihood unchanged with a fixed set of ($M_\mathrm{f},\chi_\mathrm{f}$); the likelihood is therefore only weakly informative about the calibration profile's exact shape (see Appendix~\ref{app:calib_sens} for details).

The likelihood of \cref{eq:likelihood} then generalizes to\footnote{The likelihood normalization term $-\tfrac{1}{2}\ln\det C(\{\delta\mathcal{A}_k\})$ is, strictly, no longer constant once the covariance depends on the calibration amplitude, and so does not simply drop out here as it did in \cref{eq:likelihood}. We neglect this term, since it varies negligibly across the prior, and the coverage tests below confirm this does not bias the results.}
\begin{equation}
    \label{eq:full_likelihood}
    \ln\mathcal{L}(d_t|\bm{\theta}) = -\frac{1}{2}\sum_{i,j>I_0}
    d^F_i(\bm{\theta})\,C^{-1}_{ij}(\{\delta\mathcal{A}_k\})\,d^F_j(\bm{\theta}).
\end{equation}
We place a uniform prior on $M_\mathrm{f}$ and $\chi_\mathrm{f}$ and a frequency-independent Gaussian prior on each calibration control point, with widths $\sigma_\mathcal{A}$ and $\sigma_\phi$ informed by the detector calibration uncertainty envelope
(\cref{sec:calib_unc}),
\begin{equation}
    \label{eq:prior}
    \ln\pi(\bm{\theta}) = \ln\pi(M_\mathrm{f},\chi_\mathrm{f})
    - \frac{1}{2}\sum_{k=1}^{K}\left[\frac{\delta\mathcal{A}_k^2}{\sigma_\mathcal{A}^2}
    + \frac{\delta\phi_k^2}{\sigma_\phi^2}\right] + {\rm const}.
\end{equation}
The posterior we sample is
$\ln P(\bm{\theta}|d_t) = \ln\mathcal{L}(d_t|\bm{\theta}) + \ln\pi(\bm{\theta})$,
which we explore with an ensemble Markov chain Monte Carlo (MCMC) sampler
\cite{emcee}.\footnote{For details on sampler settings see Appendix~\ref{app:MCMC}.}

Calibration marginalization need not be applied in every standard QNMRF ringdown analysis, as the additional calibration parameters increase the computational cost and increase the overall uncertainty of ($M_\mathrm{f},\chi_\mathrm{f}$) estimates. The simplest appropriate approach should be adopted based on what is known about an event's calibration:
\begin{itemize}
    \item[(a)] \textbf{Accurate calibration.} When the detector calibration uncertainty is known to be negligible within the relevant SNR regime~\cite{Sinha2025}, we can use the standard QNMRF Grid approach described in Sec.~\ref{ssec:qnmrf}.
    \item[(b)] \textbf{Well-measured nonnegligible calibration error.} When the calibration error is well constrained, either from standard detector calibration measurements~\cite{Sun2020_calib,Wade2025} or inferred from a full IMR analysis of the astrophysical signal \cite{astro_calib_paper}, we apply the MAP estimate of $\eta$ to the strain before performing the QNMRF Grid analysis as described in \cref{ssec:calib_corr}.
    \item[(c)] \textbf{Poorly constrained calibration errors.} This is the regime in which the approach described in Sec.~\ref{ssec:joint_analysis} is most suitable. When the calibration error is unknown, or poorly constrained, it is marginalized over jointly with the remnant mass and spin.
\end{itemize}
Treatments (b) and (c) are introduced in this work; (a) is the existing standard use of QNMRF. Ref.~\cite{Sinha2025} characterizes the regimes in which calibration errors become large enough to bias remnant recovery, providing practical guidance for distinguishing (a) from (b)/(c).

\section{\label{sec:sims}Simulations}

We first validate our method on simulated data.
To create realistic but controllable simulations we use an NR waveform from the Simulating eXtreme Spacetimes (SXS) catalog \cite{ SXSCatalogData_3.0.0, SXSCatalogPaper_3, SXSPackage_v2025.0.24}. Specifically, we use SXS:BBH:0305 \cite{SXS:BBH:0305}, a GW150914-like BBH merger with a remnant mass of $\Mdet=68.55\,\Msun$ and dimensionless spin of $\chi_\mathrm{f}=0.69$ (no redshift is applied, so the source- and detector-frame masses coincide here). We take only the $h_{2,\pm2}$ harmonics and set the inclination angle $\iota=0$, azimuthal angle $\beta=0$, right ascension $\alpha=0$, declination $\delta=0$ and polarization angle $\psi=0$. We take a $4\,\mathrm{s}$ signal segment with a sampling frequency of $4096\,$Hz.
The waveform is then projected onto the antenna response of the LIGO Hanford detector (H1) and injected into simulated detector noise at the level of the Advanced LIGO design sensitivity~\cite{aLIGODesign}.\footnote{Verifications of recovery of the remnant mass, spin and calibration parameters are also conducted using NR waveforms injected into white noise.} 

For testing purposes, we simulate calibration errors of the form~\cite{Sinha2025} 
\begin{align}
     \label{eq:inject_er}
    \delta\mathcal{A}(f)=\delta\mathcal{A}_{\rm{peak}}\exp\left[-\frac{1}{2}\left(\frac{f-f_{\rm{peak}}}{f_{\rm{width}}/4}\right)^2\right],\\
    \delta\phi(f)=\delta\phi_{\rm{peak}}\exp\left[-\frac{1}{2}\left(\frac{f-f_{\rm{peak}}}{f_{\rm{width}}/4}\right)^2\right].
\end{align}
This produces Gaussian-shaped errors centered on frequency $f_{\rm{peak}}$ with width $f_{\rm{width}}$ \cite{Sinha2025}. The center frequency is set to the 220 mode fundamental frequency, $f_{\rm{peak}}=249\,\textrm{Hz}$ and the width is set to $f_{\rm{width}}=50\,\textrm{Hz}$. 
This form captures the localized, narrow-band features commonly seen in real detector systematic error~\cite{Sun2020_calib, Sun2021_calib}. Following
Ref.~\cite{Sinha2025}, we simulate one such feature at a time: errors far from
the QNM frequencies have little impact on the recovered remnant, and the
Gaussian profile avoids introducing artifacts into the time-domain data.
We simulate calibration errors large enough to have a nonnegligible impact on the ringdown signal \cite{Sinha2025}. We test a range of amplitude and phase errors, including the case of no simulated calibration error. In this section, three representative examples are shown: negative amplitude and phase errors ($\delta\mathcal{A}_{\rm peak}=-0.25$, $\delta\phi_{\rm peak}=-25^\circ$; \cref{sssec:NR-neg}), positive amplitude and phase errors ($\delta\mathcal{A}_{\rm peak}=+0.25$, $\delta\phi_{\rm peak}=+25^\circ$; \cref{sssec:NR-pos}), and no calibration error ($\delta\mathcal{A}_{\rm peak}=0$, $\delta\phi_{\rm peak}=0^\circ$; \cref{sssec:NR-noerr}). 

For these examples, the noise level is set so that the injected waveform has a full-signal matched-filter SNR of $\sim60$ in one LIGO detector alone, corresponding to a ringdown-segment SNR of $\sim9$ starting at $t_0=16\,\Mdet$.\footnote{Start times are conventionally quoted in units of the remnant mass in the detector frame.}
The remnant mass and spin are recovered using a \{220, 221\} model with a $0.2\,\mathrm{s}$ data segment starting at $t_0=16\,\Mdet$. Several additional combinations of calibration errors and ringdown models were also tested but are not shown here.
The calibration control points (\cref{ssec:joint_analysis}) are evenly spaced between $\sim193.9\,$Hz and $\sim299.4\,$Hz, bracketing the 220 ($\sim249.4\,$Hz) and 221 ($\sim243.9\,$Hz) mode frequencies, and interpolated with a quadratic spline. The statistical coverage of the recovered credible intervals is assessed separately in \cref{ssec:ppplots}, over an ensemble of injections rather than the individual examples shown in \cref{sec:sims_individual}.

\subsection{Injection studies with simulated calibration errors}
\label{sec:sims_individual}
\subsubsection{Example 1: Negative error, $\delta\mathcal{A}_{\rm peak}=-0.25, \delta \phi_{\rm peak} = -25^\circ$}
\label{sssec:NR-neg}

\begin{figure}
	\includegraphics[width=\columnwidth]{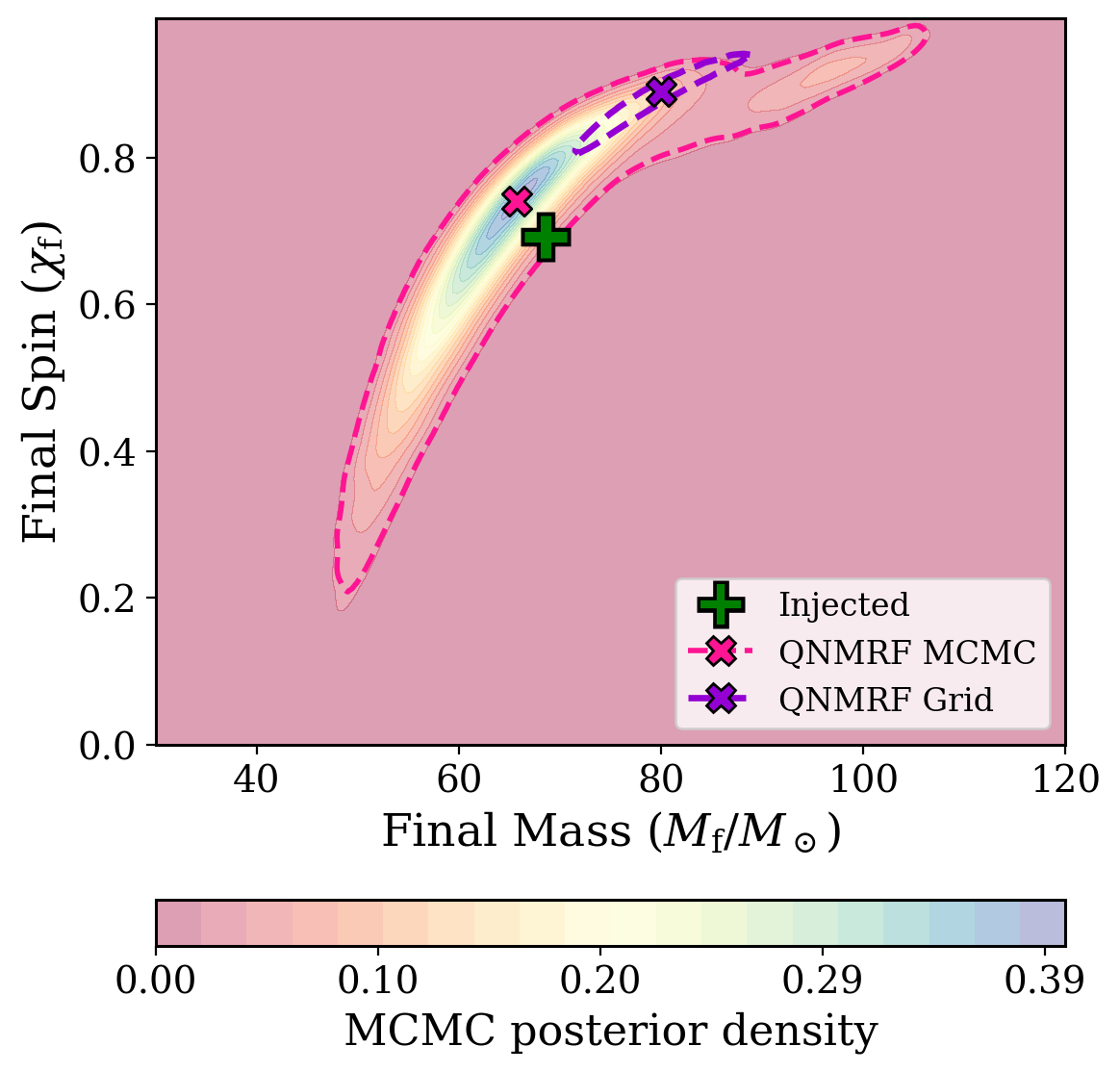}
    \caption{Mass and spin recovery for the NR waveform SXS:BBH:0305, injected with a simulated, Gaussian-shaped calibration error of $\delta\mathcal{A}_{\rm peak}=-0.25$ and $\delta\phi_{\rm peak}=-25^\circ$ centered at the fundamental-mode frequency ($f_{\rm peak}=249\,$Hz, $f_{\rm width}=50\,$Hz). The color map shows the posterior density of the QNMRF MCMC analysis, which samples over and marginalizes the calibration uncertainty using the pipeline described in Sec.~\ref{ssec:joint_analysis}. Its MAP estimate is marked by the pink cross and its $90\%$ credible region by the pink dashed contour. The green cross marks the injected (true) mass and spin. For comparison, the purple cross and dashed contour show the maximum-likelihood estimate and $90\%$ credible region of the QNMRF Grid analysis, which assumes no calibration error. The Grid analysis is biased and excludes the true values, whereas the calibration-marginalized posterior recovers them within its $90\%$ credible region.}
    \label{fig:NR_mchi_neg}
\end{figure}

The first test injects a $-25\%$ amplitude and $-25^\circ$ phase calibration error, both peaked at the fundamental $220$-mode frequency, to test whether marginalizing over calibration recovers the true remnant mass and spin that an analysis that assumes accurate calibration does not.
As shown in \cref{fig:NR_mchi_neg}, when assuming calibration is correct, the recovered mass and spin are biased high relative to the injected values (purple), with the true value lying outside the $90\%$ credible interval (CI). By contrast, applying our pipeline to the same data and jointly sampling over the calibration amplitude and phase recovers the true mass and spin within the $90\%$ CI (pink). The resulting interval is broader than that obtained in the fixed-calibration QNMRF Grid analysis, reflecting the additional uncertainty introduced by marginalizing over the calibration parameters.

The recovered calibration error for the same data is shown in \cref{fig:NR_cal_neg} (a corner plot showing all sampled parameters is provided in Appendix~\ref{app:corner_plot}). At each calibration control point, we adopt a Gaussian prior with standard deviation $20\%$ in amplitude and $20^\circ$ in phase (gray band; frequency-independent); the simulated calibration error is shown in blue. The MAP calibration error is shown as the solid pink curve, with the 68\% and 90\% CIs shaded.

The pink curve shows the calibration response evaluated at the single MCMC sample with the highest joint posterior probability, i.e., the same draw that defines the $(M_\mathrm{f}, \chi_\mathrm{f})$ MAP estimate. In amplitude, the MAP sample stays below zero across the analysis band, matching the sign of the injection, but it does not reproduce the single broad dip of the injected curve and instead exhibits additional structure.
Nevertheless, the injected amplitude error lies within the 90\% CI throughout the analysis band. 
As discussed in \cref{ssec:joint_analysis}, the likelihood carries relatively little information about the detailed frequency dependence of the calibration error, so the recovered profile is not expected to track the injected error point-by-point.  The phase error is more weakly constrained. The MAP sample recovers the correct (negative) sign only over $\sim210$--$245\,$Hz, with a shallower minimum ($\sim-10^\circ$) than the injected value ($\sim-25^\circ$). At higher frequencies, it rises to $\sim+23^\circ$ while the simulated phase error remains negative, although the injected profile lies within, or near the lower edge of the 90\% CI. Crucially, this limited recovery of the detailed frequency dependence in the calibration error profile does not compromise the remnant mass and spin estimates, consistent with the aim of the method: to marginalize over calibration uncertainty rather than to measure the calibration error.

\begin{figure}
	\includegraphics[width=\columnwidth]{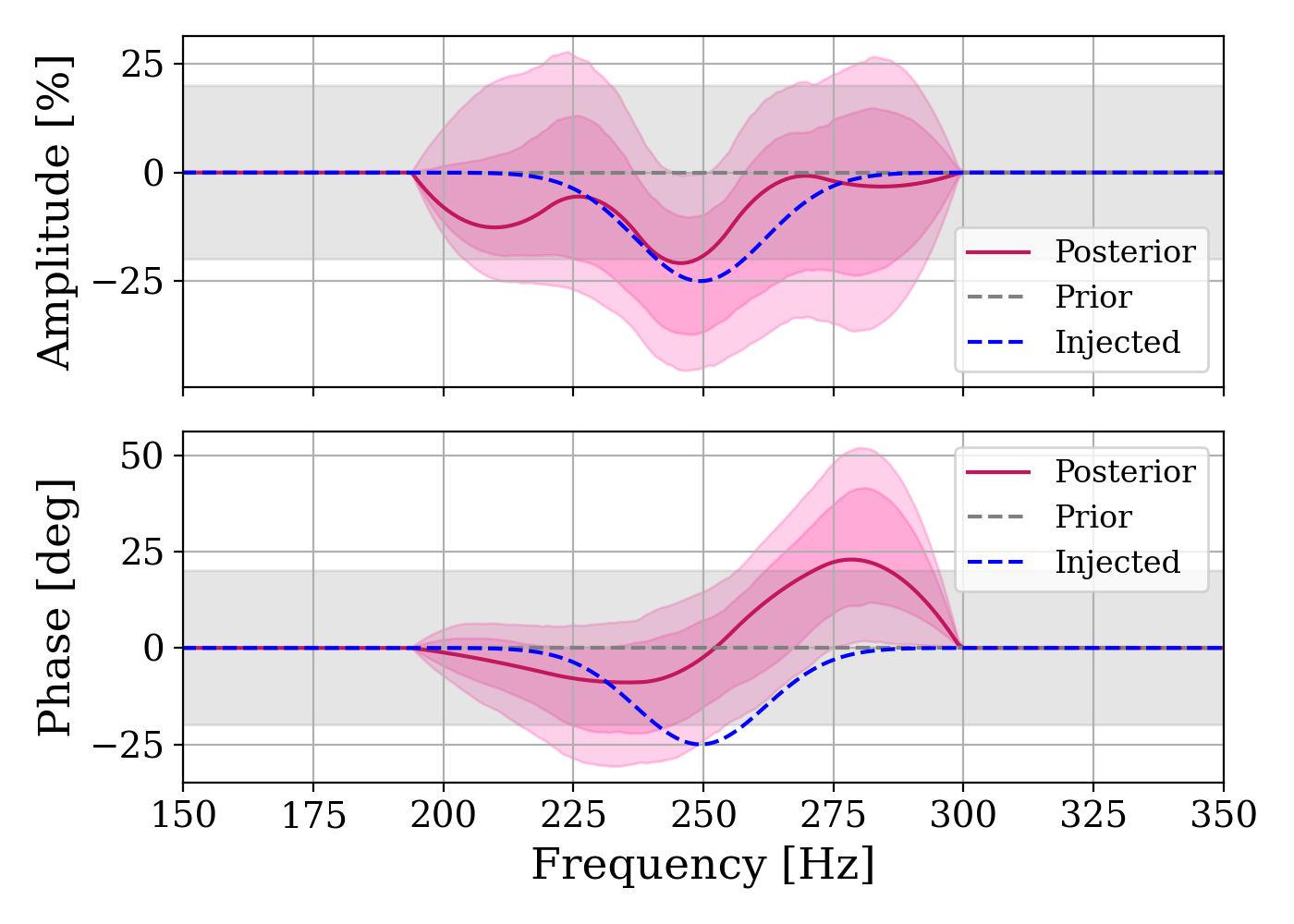}
    \caption{Recovered calibration error for SXS:BBH:0305 with an injected calibration error of $\delta\mathcal{A}_{\rm{peak}}=-0.25$ and $\delta\phi_{\rm{peak}}=-25^\circ$ (a Gaussian-shaped error centered at $f_{\rm peak}=249\,$Hz with $f_{\rm width}=50\,$Hz). The injected error is shown in blue. A frequency-independent Gaussian calibration-error prior (gray), with $1\sigma$ width $20\%$ in amplitude and $20^\circ$ in phase, is used at each control point. The posterior from the MCMC analysis is shown in pink. The MAP calibration error is shown in dark pink, with the $68\%$ (light pink) CI and $90\%$ CI (lighter pink) shaded.}
    \label{fig:NR_cal_neg}
\end{figure}

\subsubsection{Example 2: Positive error, $\delta\mathcal{A}_{\rm peak}=+0.25, \delta \phi_{\rm peak} = +25^\circ$}
\label{sssec:NR-pos}

\begin{figure}
	\includegraphics[width=\columnwidth]{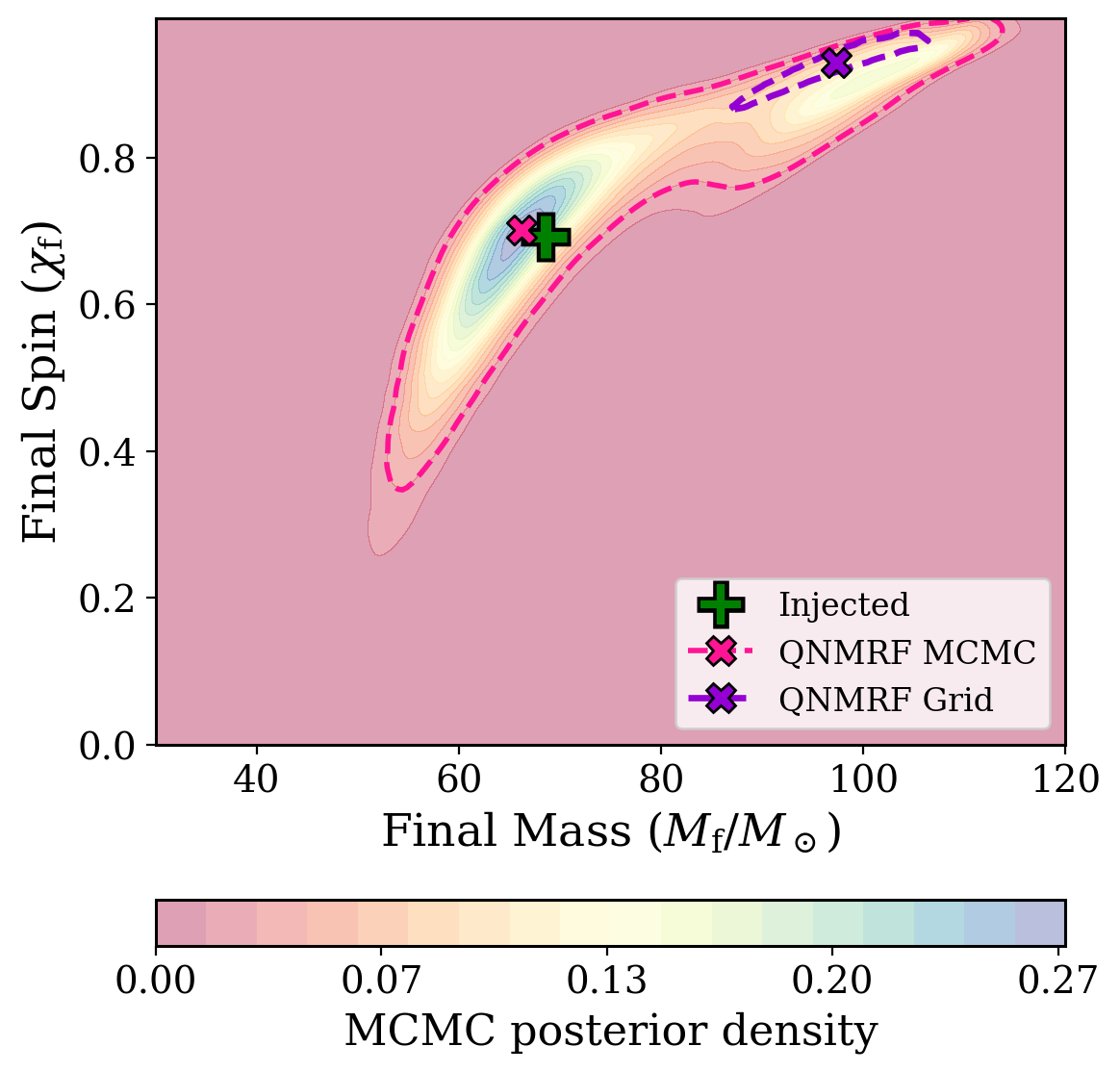}
    \caption{As \cref{fig:NR_mchi_neg} but with a simulated calibration error of $\delta\mathcal{A}_{\rm peak}=+0.25$ and $\delta\phi_{\rm peak}=+25^\circ$ (centered at $f_{\rm peak}=249\,$Hz with $f_{\rm width}=50\,$Hz).}
    \label{fig:NR_mchi_pos}
\end{figure}

To demonstrate that the pipeline is robust to the direction of the calibration error, we repeat the analysis with a positive injected error here ($\delta\mathcal{A}_{\rm{peak}}=0.25$ and $\delta\phi_{\rm{peak}}=+25^\circ$), opposite in sign to the one tested in \cref{sssec:NR-neg}. We once again see in \cref{fig:NR_mchi_pos} that the Grid solution (purple) is biased when neglecting the calibration error, with the true value lying outside the 90\% CI. The calibration-marginalized posterior (pink) recovers the true value.  

\begin{figure}
	\includegraphics[width=\columnwidth]{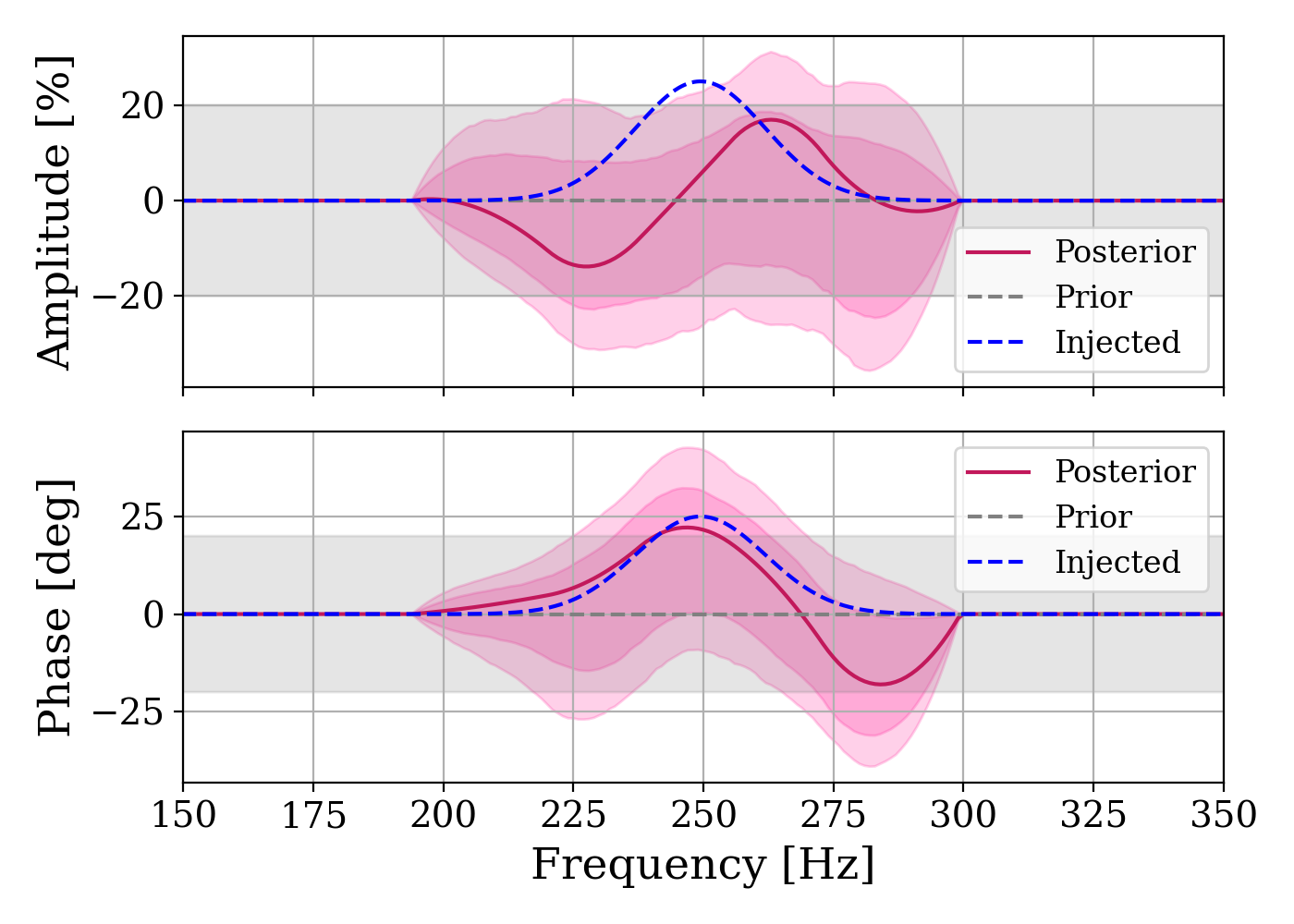}
    \caption{As \cref{fig:NR_cal_neg} but with a simulated calibration error of $\delta\mathcal{A}_{\rm peak}=+0.25$ and $\delta\phi_{\rm peak}=+25^\circ$ (centered at $f_{\rm peak}=249\,$Hz with $f_{\rm width}=50\,$Hz).}
    \label{fig:NR_cal_pos}
\end{figure}

The recovered amplitude error in \cref{fig:NR_cal_pos} again captures the sign of the injection, with a positive excursion peaking near $\sim260$--$265\,$Hz, though the recovered peak ($\sim17\%$) is lower than, and shifted in frequency from, the injected peak ($\sim25\%$ at $\sim250\,$Hz); it is also accompanied by a comparably sized negative excursion, reaching $\sim-14\%$ near $\sim225$--$230\,$Hz, not present in the injection. As discussed in \cref{ssec:joint_analysis}, such structure reflects the weak constraint that the likelihood places on the detailed calibration profile. The phase error is better recovered than the case in Sec.~\ref{sssec:NR-neg}: the MAP sample reproduces both the sign and approximate magnitude of the injected phase bump near the fundamental mode, reaching $\sim+23^\circ$ close to the injected peak, before overshooting to the opposite sign above $\sim270\,$Hz. As in all other simulations, this limited recovery of the detailed calibration profile does not degrade the mass and spin estimates.

\subsubsection{Example 3: No error, $\delta\mathcal{A}=0, \delta \phi = 0^\circ$}
\label{sssec:NR-noerr}

\begin{figure}
	\includegraphics[width=\columnwidth]{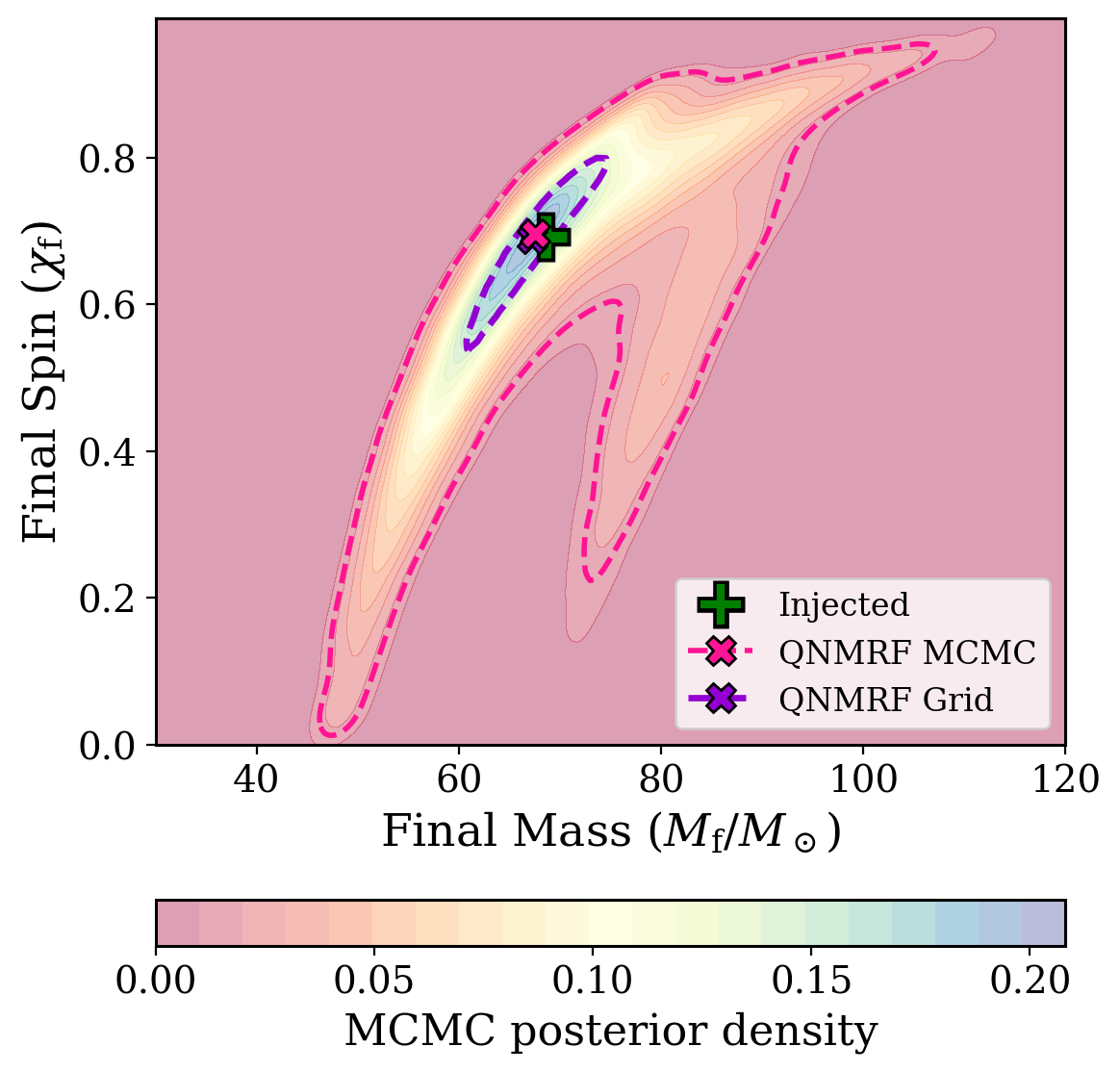}
    \caption{As \cref{fig:NR_mchi_neg} but with no calibration error injected.}
    \label{fig:NR_mchi_noerr}
\end{figure}

\begin{figure}
	\includegraphics[width=\columnwidth]{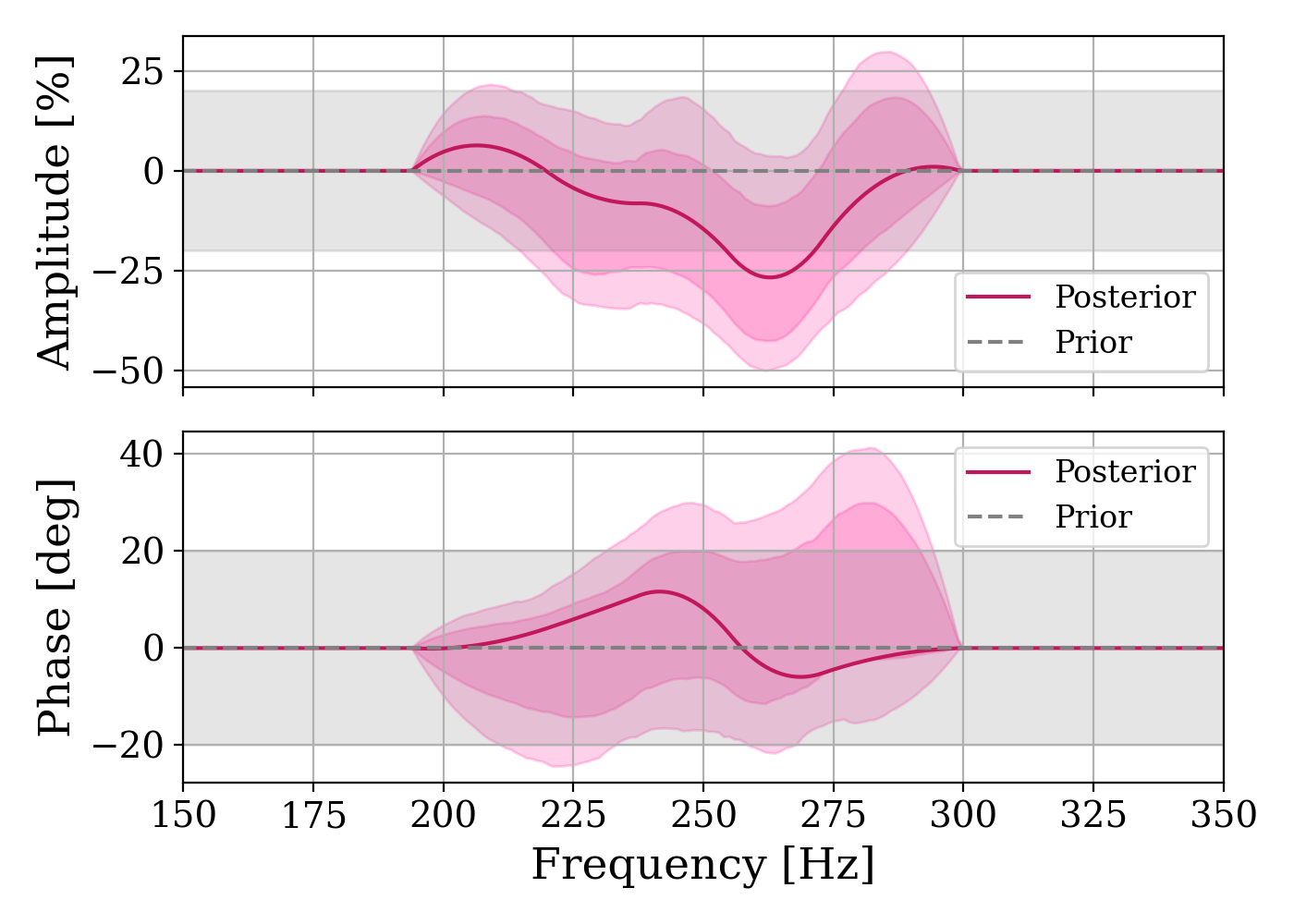}
    \caption{As \cref{fig:NR_cal_neg} but with no calibration error injected.}
    \label{fig:NR_cal_noerr}
\end{figure}

We also perform an important null test, where no calibration error is injected, to quantify the cost of marginalizing over calibration uncertainty when no error is present. We show in \cref{fig:NR_mchi_noerr} that both the QNMRF Grid (purple) and calibration-marginalized MCMC (pink) recover the true value. The cost, however, is a substantially larger $90\%$ credible region under marginalization. Thus, when no calibration error is present in the data, the calibration-marginalized MCMC pipeline is conservative, increasing the uncertainty on the estimated remnant parameters. Additionally, the recovered calibration posterior in \cref{fig:NR_cal_noerr} is consistent with zero error within the 90\% CI.

\subsection{P--P plots}
\label{ssec:ppplots}

A single noise realization, such as those shown in \cref{sec:sims}, cannot establish that our CIs are statistically meaningful, as a favorable noise draw could make the pipeline appear more accurate than it is. To more comprehensively test this, we inject an analytic $220$ damped sinusoid into many independent realizations of colored noise at the design sensitivity of Advanced LIGO \cite{aLIGODesign} and construct probability--probability (P--P) plots to assess the statistical coverage of the recovered CI. For each realization, we record the credible level at which the true mass and spin fall. Well-calibrated posteriors produce uniformly distributed credible levels, so the cumulative fraction of realizations follows the diagonal. We perform two tests, identical except for the calibration error present in the injected data.

Each realization draws a fresh set of true parameters from the same priors the recovery assumes (the remnant mass is drawn uniformly from $M_\mathrm{f}\in[45,75]\,\Msun$ and the spin uniformly from $\chi_\mathrm{f}\in[0.5,0.9]$). The calibration control points are placed at the same frequencies across all realizations.\footnote{The control points span a wider range than in \cref{sec:sims} ($213\,$Hz to $468\,$Hz) to accommodate the range of remnant masses and therefore fundamental-mode frequencies across all realizations.} The two
ensembles differ only in the calibration error injected into the data:
\begin{itemize}
    \item[(1)] \textbf{Errors drawn from the prior.} At each control point, the amplitude and phase errors are
    drawn independently from zero-mean Gaussians of width $20\%$ in amplitude and
    $20^\circ$ in phase (the same priors assumed in the recovery). The injected calibration error is then obtained by spline interpolation through these randomly drawn control points.
    \item[(2)] \textbf{No injected error.} Every control point is set to zero, corresponding to perfectly calibrated data.
\end{itemize}

The injected signal is an analytic $220$ damped sinusoid rather than the NR waveform of \cref{sec:sims_individual}. The P--P test assesses whether the inference is statistically self-consistent, which requires the data to be generated from the same model the analysis assumes. A full NR ringdown contains overtones and nonlinear content that the rational filter cannot represent exactly at any $M_\mathrm{f}$ and $\chi_\mathrm{f}$, so a departure from the diagonal could otherwise arise from waveform mismatch rather than the calibration treatment. Injecting a signal represented exactly by the analysis model removes that ambiguity, leaving the calibration treatment as the only difference between the two tests.

\begin{figure}
	\includegraphics[width=\columnwidth]{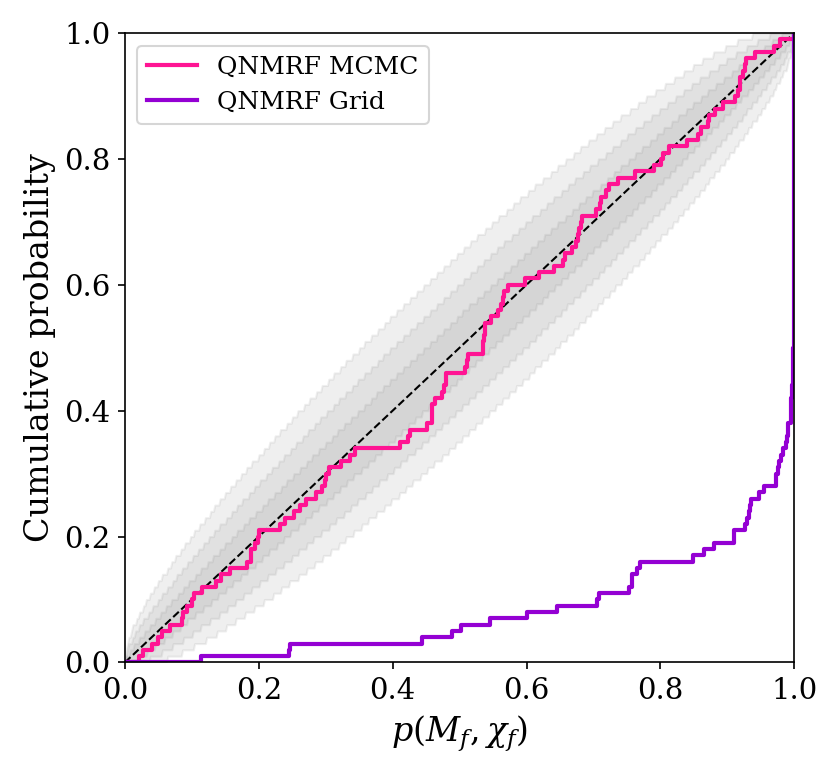}
    \caption{Joint P--P plot for remnant mass and spin over 100 injections with calibration errors drawn from the prior described by (1) above. 
    Each curve shows the cumulative fraction of runs for which the true $(M_\mathrm{f}, \chi_\mathrm{f})$ fall within a given joint credible level $p(M_\mathrm{f}, \chi_\mathrm{f})$; a well-calibrated analysis follows the diagonal (gray bands show the $1, 2$ and $3\sigma$ expected scatter). Only the pipeline that marginalizes over calibration (pink) is consistent with the diagonal; the fixed-calibration Grid method (purple) departs sharply from it, with CIs systematically excluding the true mass and spin.}
    \label{fig:pp_priordraw}
\end{figure}

When the injected calibration errors are drawn from the prior (\cref{fig:pp_priordraw}), only the calibration-marginalized analysis achieves correct coverage. Neglecting calibration error therefore does not merely broaden the recovered mass and spin; it makes the credible intervals themselves unreliable.

\begin{figure}
	\includegraphics[width=\columnwidth]{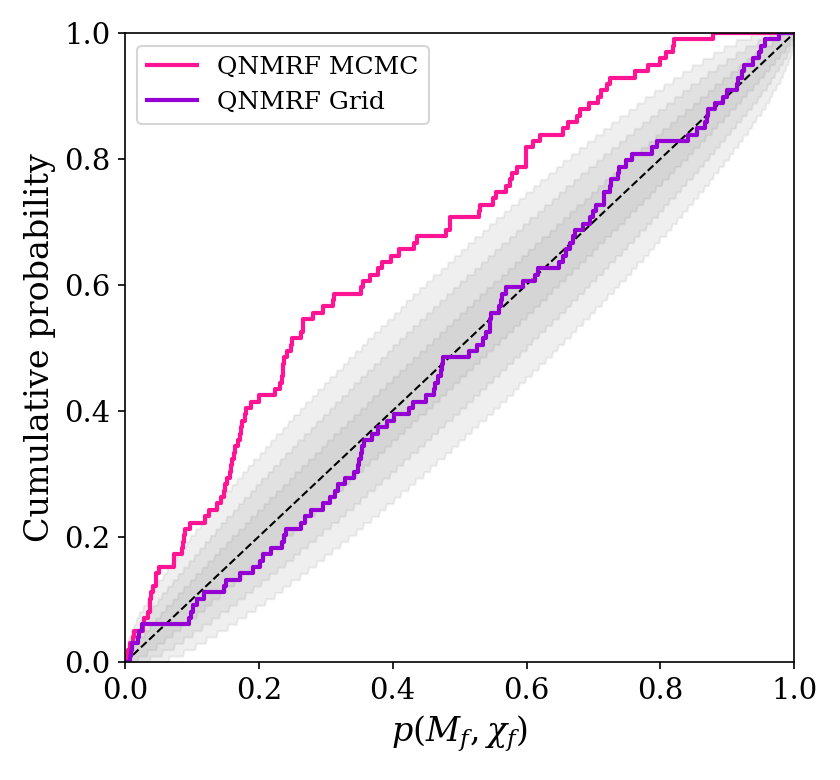}
    \caption{As \cref{fig:pp_priordraw}, but for 100 injections with no calibration error present [test (2) above]. The fixed-calibration Grid analysis (purple) traces the diagonal, since the data match the zero-error assumption. The calibration-marginalized MCMC pipeline (pink) lies above the diagonal, indicating over-coverage.} 
    \label{fig:pp_zeros}
\end{figure}

In the null case, where no calibration error is injected (\cref{fig:pp_zeros}), the fixed-calibration Grid analysis is, by construction, well calibrated. The marginalized MCMC pipeline is not expected to trace the diagonal here, since uniform credible levels are guaranteed only when the simulated calibration error is itself drawn from the assumed prior, rather than fixed at zero. The P--P curve instead lies above the diagonal, reflecting over-coverage: marginalizing over calibration uncertainty that is absent from the data conservatively inflates the credible regions, so the true parameters are enclosed more often than nominal. This is the conservative direction, in contrast to the under-coverage (a P--P curve below the diagonal) that would indicate genuinely overconfident posteriors. Together, these tests demonstrate that calibration marginalization restores correct statistical coverage in the regime where neglecting it would bias the inferred mass and spin, while becoming conservative when no calibration error is present. 

\section{\label{sec:results}Application to GW250207}

GW250207 was recorded on February 7, 2025 at 11:56:45~UTC during the third part of the LVK fourth observing run (O4c)~\cite{astro_calib_paper, gwosc_O4b, astro_calib_zenodo}. 
It was identified using the LIGO Livingston (L1) and Virgo (V1) detectors; H1 was at the beginning of a lock stretch, recording but not yet stabilized, leading to poorly constrained calibration~\cite{astro_calib_paper}. This makes GW250207 an ideal test of our pipeline: H1 strain data are available but with a large calibration uncertainty.

This event has since been calibrated \emph{astrophysically} in Ref.~\cite{astro_calib_paper}, which jointly infers the source parameters and a frequency-dependent H1 calibration-error spline across the full signal band, using the same $\eta(f)$ parameterization as \cref{eq:calib}. We denote the MAP calibration estimate 
\begin{equation}
    \label{eq:astro_calib_est}
    \eta_{\rm PE}^{\rm MAP}(f) = \big[1+\delta\mathcal{A}_{\rm PE}^{\rm MAP}(f)\big]\exp\!\big[i\,\delta\phi_{\rm PE}^{\rm MAP}(f)\big],
\end{equation}
a single spline curve, which we construct from the public posterior samples \cite{astro_calib_zenodo} released with Ref.~\cite{astro_calib_paper} by taking the draw that maximizes the joint posterior over all sampled parameters and interpolating between its calibration nodes to generate $\delta\mathcal{A}_{\rm PE}^{\rm MAP}$ and $\delta\phi_{\rm PE}^{\rm MAP}$.
Wherever a treatment of the H1 data below is labeled ``$\eta_{\rm PE}^{\rm MAP}$ correction'' (\cref{tab:250207_Mchi}, \cref{fig:250207_Mchi_2det,fig:250207_Mchi_1det,fig:250207_calib}), this
externally estimated correction has been applied to the strain following the procedure of \cref{ssec:calib_corr}.

This is distinct from the calibration-marginalized QNMRF MCMC treatment, in which $\eta(f;\bm\theta)$ of \cref{eq:calib} is left free and sampled jointly with $M_\mathrm{f}$ and $\chi_\mathrm{f}$ under the Gaussian prior of \cref{eq:prior}. In this case, the frequency-dependent calibration \emph{uncertainty}, rather than a single estimate of it, is propagated into the remnant posterior. Thus, we analyze the data \emph{without} applying $\eta_{\rm PE}^{\rm MAP}$ to assess whether our ringdown-only pipeline can recover the remnant mass and spin from poorly calibrated data. We then compare both the inferred remnant properties and calibration error against the estimates presented in Ref.~\cite{astro_calib_paper}, which provides an independent, full-signal benchmark.

It has already been shown in Ref.~\cite{astro_calib_paper} that the calibration errors in H1 induce frequency and damping-time deviations for the $220$ mode in GW250207, further motivating the need to account for these errors in our analysis. The event has a network SNR of $68.91$, corresponding to a ringdown SNR of $\sim20$. 
We consider start times $t_0$ from $0$ to $15\,\MIMR$ after the peak strain amplitude of the MAP surrogate waveform~\cite{Varma_2019_NRSur7dq4}, in integer multiples of the remnant mass ($1\,\MIMR=3.22\times10^{-4}\,\mathrm{s}$). The sky location is fixed throughout to the MAP estimate from the IMR analysis.\footnote{The strain peak time used is $t_\mathrm{peak}=1422964623.267101\,\mathrm{s}$; sky location is fixed at right ascension $\alpha=2.84$\,rad and declination $\delta=0.63$\,rad.} To identify which ringdown modes are present and the appropriate start time, we perform the analysis for both the joint H1--L1 (Sec.~\ref{ssec:res_2det}) and H1-only data (Sec.~\ref{ssec:res_1det}). 

Throughout this section, when recovering mass, spin and calibration, the MCMC analyses use the following priors: $M_\mathrm{f}\sim\mathcal{U}(32.7,98.1)\,\Msun$,\footnote{These mass values are found using $0.5\MIMR$ and $1.5\MIMR$.} $\chi_\mathrm{f}\sim\mathcal{U}(0,0.99), \delta\mathcal{A}_k\sim\mathcal{N}(0,\sigma_\mathcal{A}^2)$ and
$\delta\phi_k\sim\mathcal{N}(0,\sigma_\phi^2)$ for $k=1,\dots,5$, with
$\sigma_\mathcal{A}=0.20$ and $\sigma_\phi=20^\circ$. These widths are
deliberately broader than the O4 state-of-the-art quoted in \cref{sec:calib_unc}, since the H1 response was not well characterized at the time of this event. The Grid analyses evaluate the likelihood over the same mass and spin range. The calibration of L1 is held fixed throughout.

\subsection{Hanford--Livingston analysis}
\label{ssec:res_2det}

We first apply our pipeline to the joint H1--L1 data, sweeping over the ringdown start time $t_0$ to identify which QNMs are present and assess the recovery of remnant mass and spin, using the consistency with the IMR prediction as a diagnostic. Having fixed a start time and mode content, we then compare the mass, spin and calibration recovery under three calibration treatments: leaving the error uncorrected, applying the $\eta_{\rm PE}^{\rm MAP}$ correction, and marginalizing over the error.

\begin{figure}
    \includegraphics[width=\columnwidth]{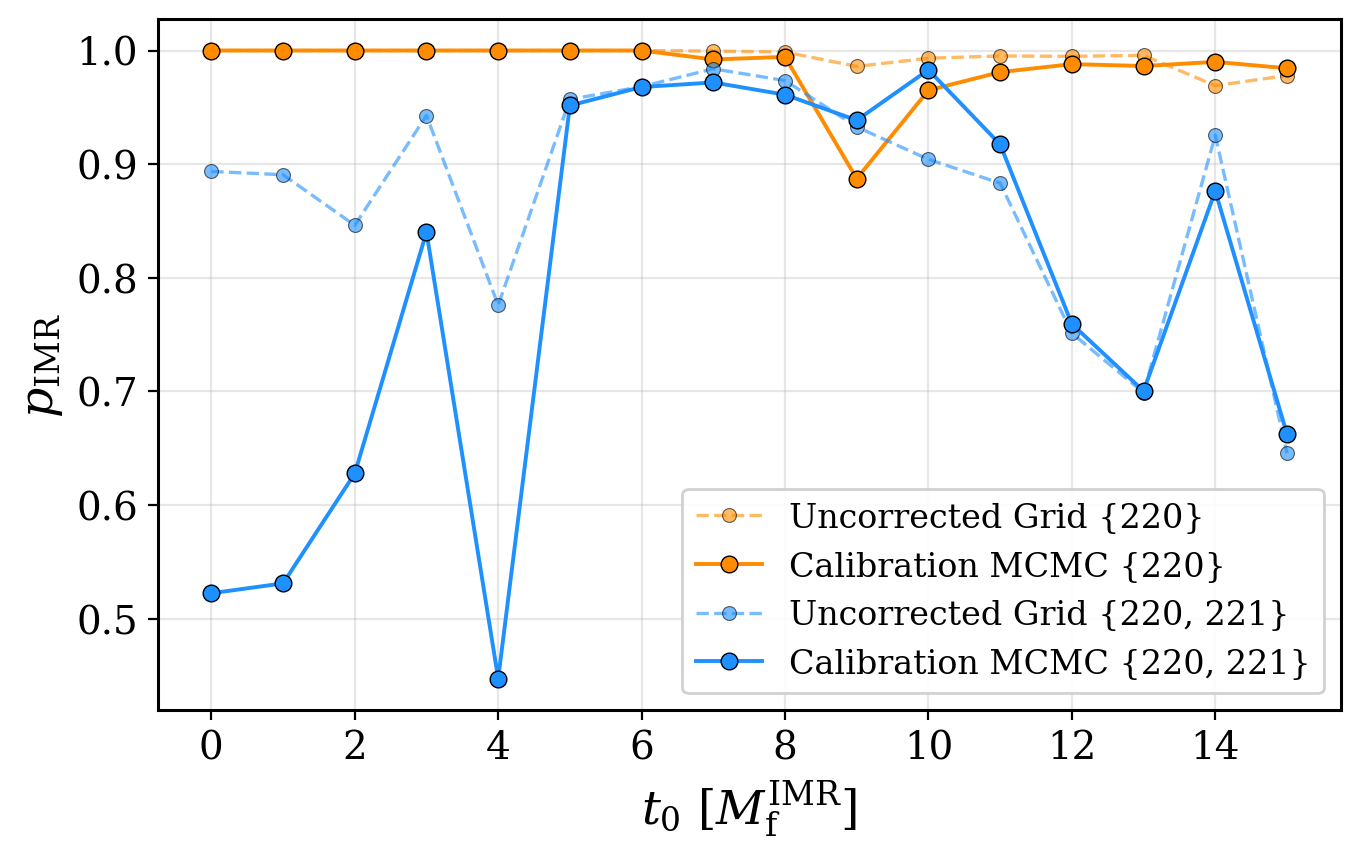}
    \caption{Joint posterior quantile $p_\mathrm{IMR}$ as a function of the start time $t_0$ after the strain peak (in units of the remnant mass $\MIMR$) for GW250207 using the two-detector H1--L1 data. Orange and blue indicate the $\{220\}$-only and $\{220,221\}$ models, respectively. Dashed curves show results with the H1 calibration error uncorrected, while solid curves show results obtained by marginalizing over the H1 calibration uncertainty. L1 data has not had any correction or calibration marginalization applied. After $\sim 10\,\MIMR$, the data become noise dominated, so any improvement in $p_{\rm IMR}$ reflects broadening of the CIs rather than a genuine improvement in the estimate.}
    \label{fig:250207_2det}
\end{figure}

In \cref{fig:250207_2det}, we quantify the consistency of the remnant mass and spin inferred from the ringdown data with the IMR prediction using $p_{\rm IMR}$ (a joint posterior quantile where the IMR-predicted values lie), defined as~\cite{Lu2025_rf}
\begin{equation}
    \label{eq:pIMR}
    p_\mathrm{IMR}=\frac{\sum_{\mathcal{L}(d|\Mdet,\chi_\mathrm{f})>\mathcal{L}(d|\Mdet^\mathrm{IMR},\chi_\mathrm{f}^\mathrm{IMR})}\mathcal{L}(d|\Mdet,\chi_\mathrm{f})}{\sum_{\Mdet,\chi_\mathrm{f}}\mathcal{L}(d|\Mdet,\chi_\mathrm{f})}.
\end{equation}
A value below $0.9$ places the IMR value inside the $90\%$ credible region, while smaller values indicate closer agreement with the posterior peak. The fundamental-only $\{220\}$ analysis remains in tension with the IMR remnant across most early start times ($p_{\rm IMR}\!\approx\!1$), whereas including the 221 overtone brings the recovery into agreement at early times. Calibration marginalization (QNMRF MCMC) further improves the agreement of the $\{220, 221\}$ model with the IMR prediction at early times. Beyond  $\sim 10\,\MIMR$ the signal is dominated by noise; consequently, the apparent decrease in $p_{\rm IMR}$ at later times arises from the CIs broadening, rather than from any genuine improvement in mass and spin estimate.\footnote{Standard QNMRF analyses also evaluate a detection statistic across start time to compare mode hypotheses~\cite{Lu2025_rf}. Robust interpretation requires a noise-background study to exclude noise mimicking an additional QNM, which we defer to future work.}

\begin{figure}
    \includegraphics[width=\columnwidth]{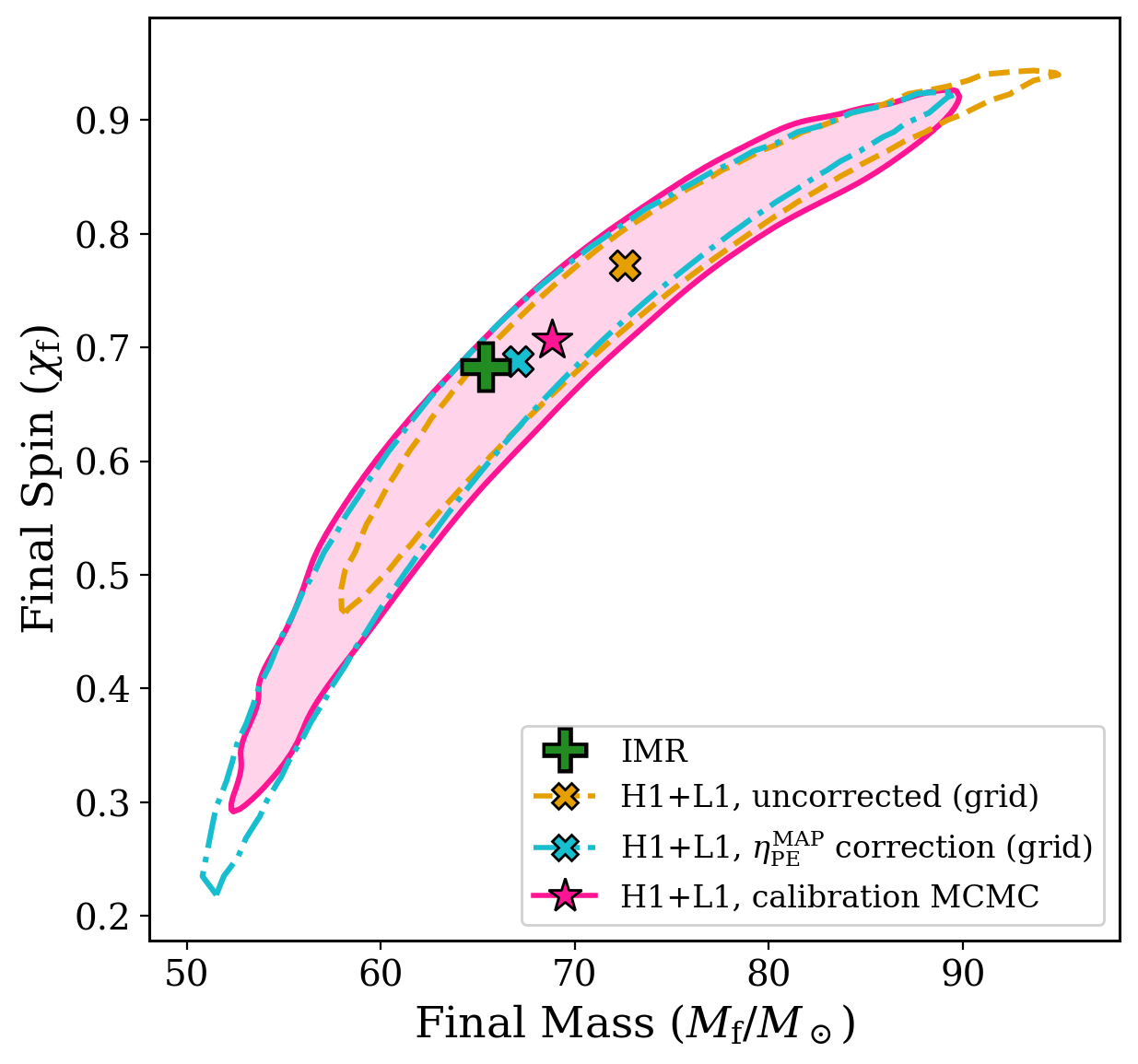}
    \caption{Remnant mass ($M_\mathrm{f}$) and spin ($\chi_\mathrm{f}$) recovery for the two-detector (H1--L1) GW250207 analysis under different calibration treatments, using a \{220, 221\} model at $t_0=4\,\Mdet^\mathrm{IMR}$. The green plus marks the remnant predicted by the full IMR analysis. All three treatments use the QNMRF pipeline of \cref{sec:filter_MCMC}: H1 calibration left uncorrected (yellow), H1 corrected using $\eta_{\rm PE}^{\rm MAP}$ from \cite{astro_calib_paper} (cyan), and H1 calibration marginalized (pink). For L1, we use the standard QNMRF Grid method without accounting for calibration. Crosses denote the maximum-likelihood estimates from the Grid analyses, while the star marks the MAP estimate from the MCMC analysis. The dashed, dot-dashed, and solid contours enclose the corresponding $90\%$ credible regions.}
    \label{fig:250207_Mchi_2det}
\end{figure}

Guided by \cref{fig:250207_2det}, we adopt the $\{220, 221\}$ model at $t_0 = 4\,\MIMR$, where the recovered mass and spin are most consistent with the IMR prediction, for the detailed mass--spin and
calibration recovery presented in the remainder of this section
(\cref{tab:250207_Mchi}, \cref{fig:250207_Mchi_2det}). The calibration control points (\cref{ssec:joint_analysis}) are evenly spaced between $\sim203.6\,$Hz and $\sim309.6\,$Hz, spanning both mode frequencies.

The remnant mass and spin recovery is shown in \cref{fig:250207_Mchi_2det}, with the MAP estimate and $90\%$ credible region of the calibration-marginalized analysis (QNMRF MCMC) shown in pink. Results for the different calibration treatments are compared using the joint posterior quantile $p_{\rm IMR}$ in \cref{tab:250207_Mchi}. 

Including the H1 data without calibration correction biases the recovered maximum-likelihood remnant properties; applying the correction factor $\eta_{\rm PE}^{\rm MAP}$ or marginalizing the calibration uncertainty reduces the maximum-likelihood/MAP point estimate discrepancy by more than a factor of two.
The QNMRF MCMC analysis yields the $p_{\rm IMR}$ most consistent with the IMR prediction, comparable to applying the correction factor $\eta_{\rm PE}^{\rm MAP}$ inferred from the full signal.

\begin{table}
    \centering
    \begin{tabular}{|l|l|c|c|c|}
        \hline
        Data & Calibration treatment & $M_\mathrm{f}$ & $\chi_\mathrm{f}$ & $p_{\rm IMR}$ \\
        \hline
        IMR (full signal) & --          & 65.38 & 0.683 & -- \\
        H1+L1             & uncorrected & 72.59 & 0.772 & 0.776 \\
        H1+L1             & $\eta_{\rm PE}^{\rm MAP}$ correction & 67.04 & 0.688 & 0.477  \\
        H1+L1             & calibration MCMC & 68.83 & 0.706 & 0.444 \\
        \hline
    \end{tabular}
        \caption{Maximum-likelihood (Grid) or MAP (MCMC) remnant $M_\mathrm{f}$ and $\chi_\mathrm{f}$ estimates for GW250207 ($t_0=4\,\MIMR$, filters $\{220,221\}$) under each calibration treatment, together with the joint posterior quantile $p_\mathrm{IMR}$ of the IMR-predicted remnant, see \cref{eq:pIMR}.}
    \label{tab:250207_Mchi}
\end{table}

\begin{figure}
    \includegraphics[width=\columnwidth]{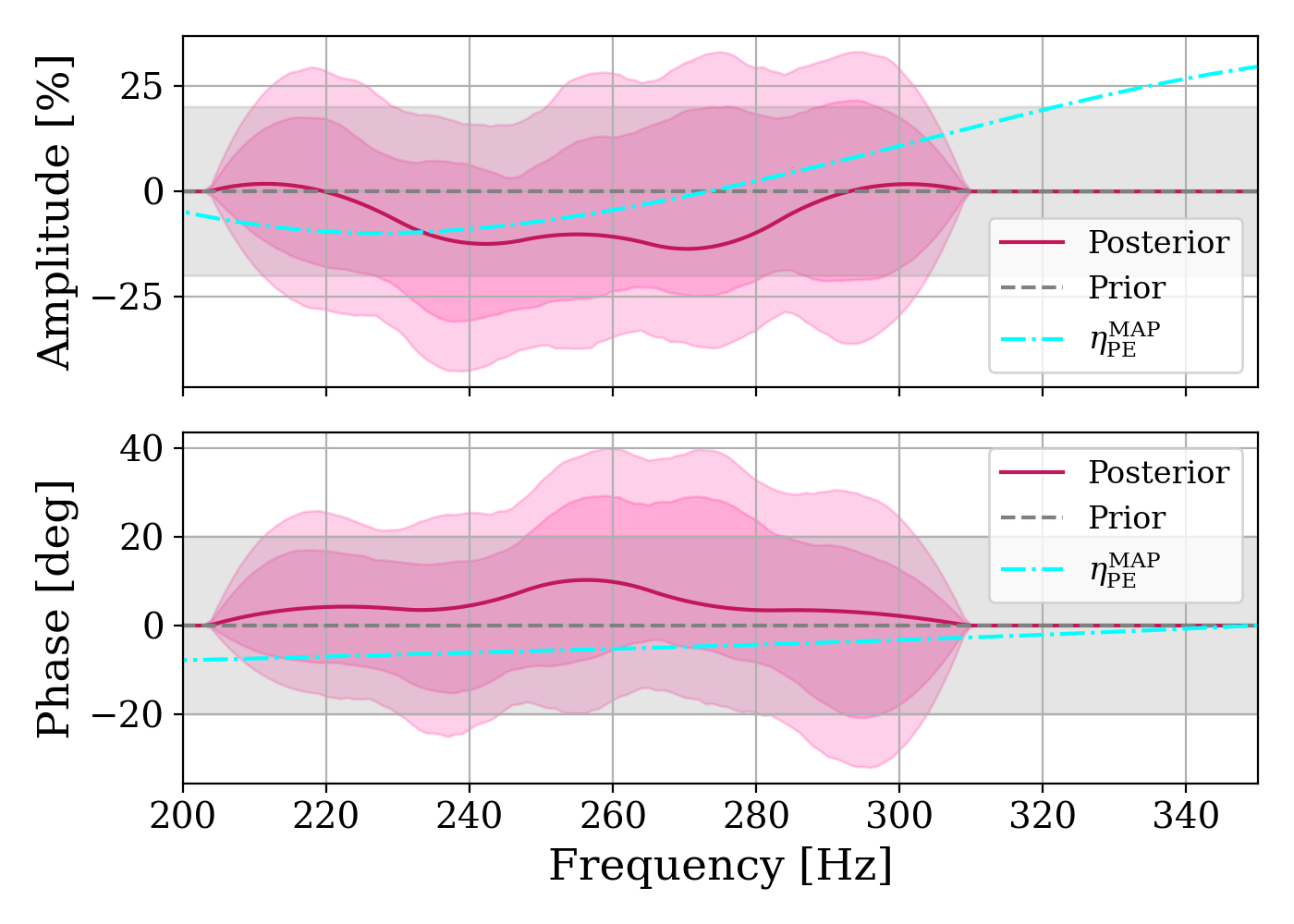}
    \caption{Recovered H1 calibration error for GW250207 at $ t_0=4\,\MIMR$ using the $\{220, 221\}$ model in the two-detector analysis. The QNMRF MCMC posterior (pink, with solid curve showing the MAP estimate and the shaded region indicating the $68\%$ and $90\%$ credible bands) is compared with $\eta_{\rm PE}^{\rm MAP}$ from \cref{eq:astro_calib_est},
    inferred from the full signal in Ref.~\cite{astro_calib_paper} (cyan). The frequency-independent Gaussian prior with $1\sigma$ width of $20\%$ in amplitude and $20^\circ$ in phase is shown in gray. The recovery is only relevant within the ringdown analysis band ($\sim$204--310~Hz), and is set to zero elsewhere.}
    \label{fig:250207_calib}
\end{figure}

The H1 calibration error recovered is shown in \cref{fig:250207_calib}. Despite the moderate ringdown SNR, the posterior shows a mild departure from the prior within the analysis band. In amplitude, the MAP curve forms a broad trough reaching $\sim-13\%$ across $\sim230$--$280\,$Hz, and in phase a broad excess peaking at $\sim+10^\circ$ near $255\,$Hz, with both returning towards the prior at the edges of the band. This departure is not statistically significant, however, as zero error lies within the $68\%$ CI throughout. Therefore, the ringdown shows at most mild sensitivity to the H1 miscalibration and does not recover the detailed calibration error profile $\eta_{\rm PE}^{\rm MAP}$ of Ref.~\cite{astro_calib_paper}.
Nonetheless, $\eta_{\rm PE}^{\rm MAP}$ remains statistically consistent with our ringdown-only recovery in the relevant band, lying within the $68\%$ CI in both amplitude and phase. Given the ringdown SNR of only $\sim20$, such a weakly constrained calibration recovery is not surprising.

\subsection{Hanford-only analysis}
\label{ssec:res_1det}

\begin{figure}
    \includegraphics[width=\columnwidth]{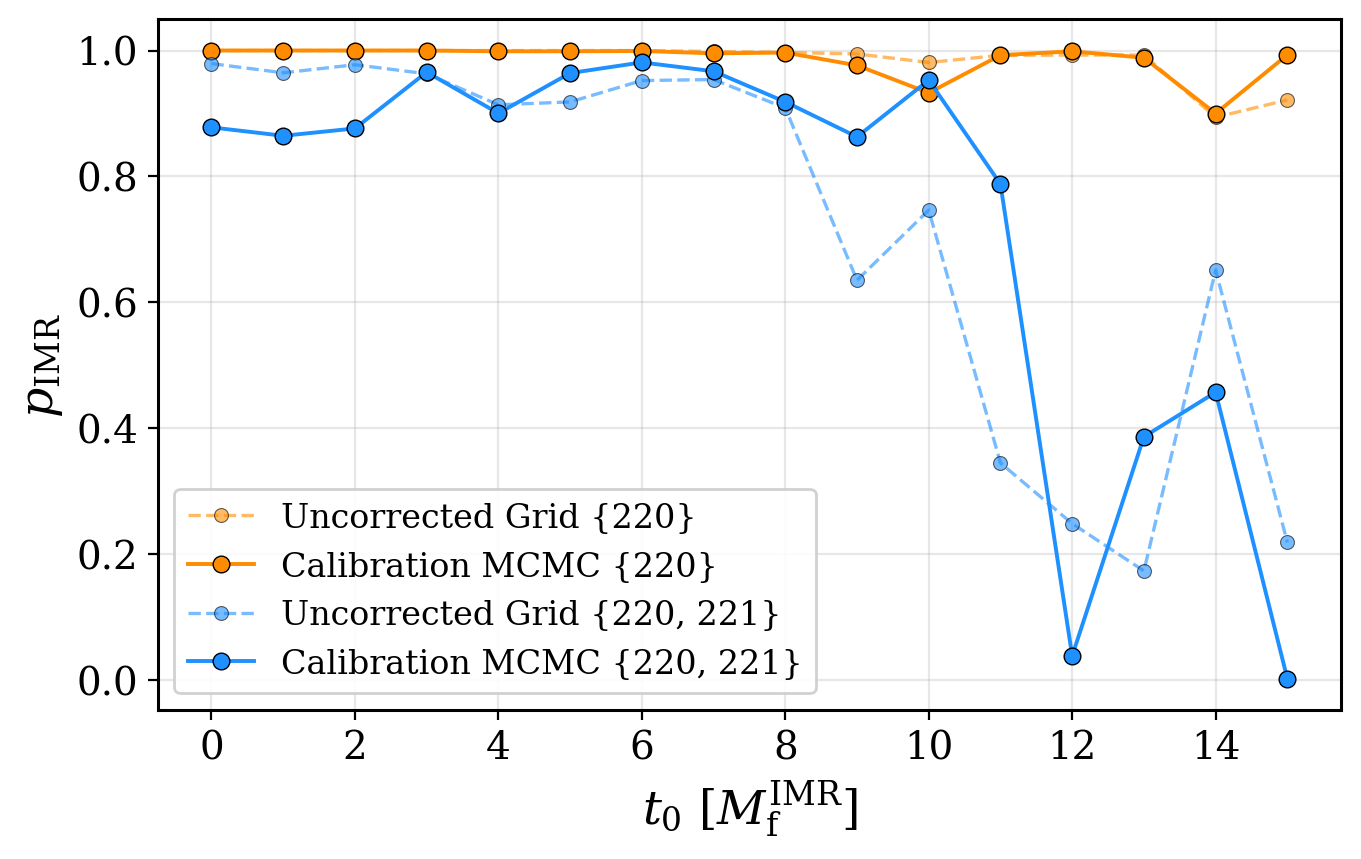}
    \caption{As \cref{fig:250207_2det}, but for the H1 detector only. }
    \label{fig:250207_1det}
\end{figure}

We perform the same analysis but for H1-only data. The agreement with the IMR mass and spin is shown in \cref{fig:250207_1det}. Once again, the dashed curves denote results from ignoring calibration errors, and the solid curves indicate results obtained by marginalizing over the calibration uncertainty. The agreement with the IMR value for both mode models is generally poor, and the calibration marginalization yields little improvement. This is likely due to the absence of a second well-calibrated detector that can constrain $M_\mathrm{f}$ and $\chi_\mathrm{f}$ independently of the H1 data, and to the lower SNR of single-detector data. Similar to the two-detector case, the occasional dips to low $p_{\rm IMR}$ at later start times (e.g. $t_0=12\,\MIMR$ and $15\,\MIMR$) likely reflect this loss of constraining power rather than genuine improvement.

\begin{figure}
    \includegraphics[width=\columnwidth]{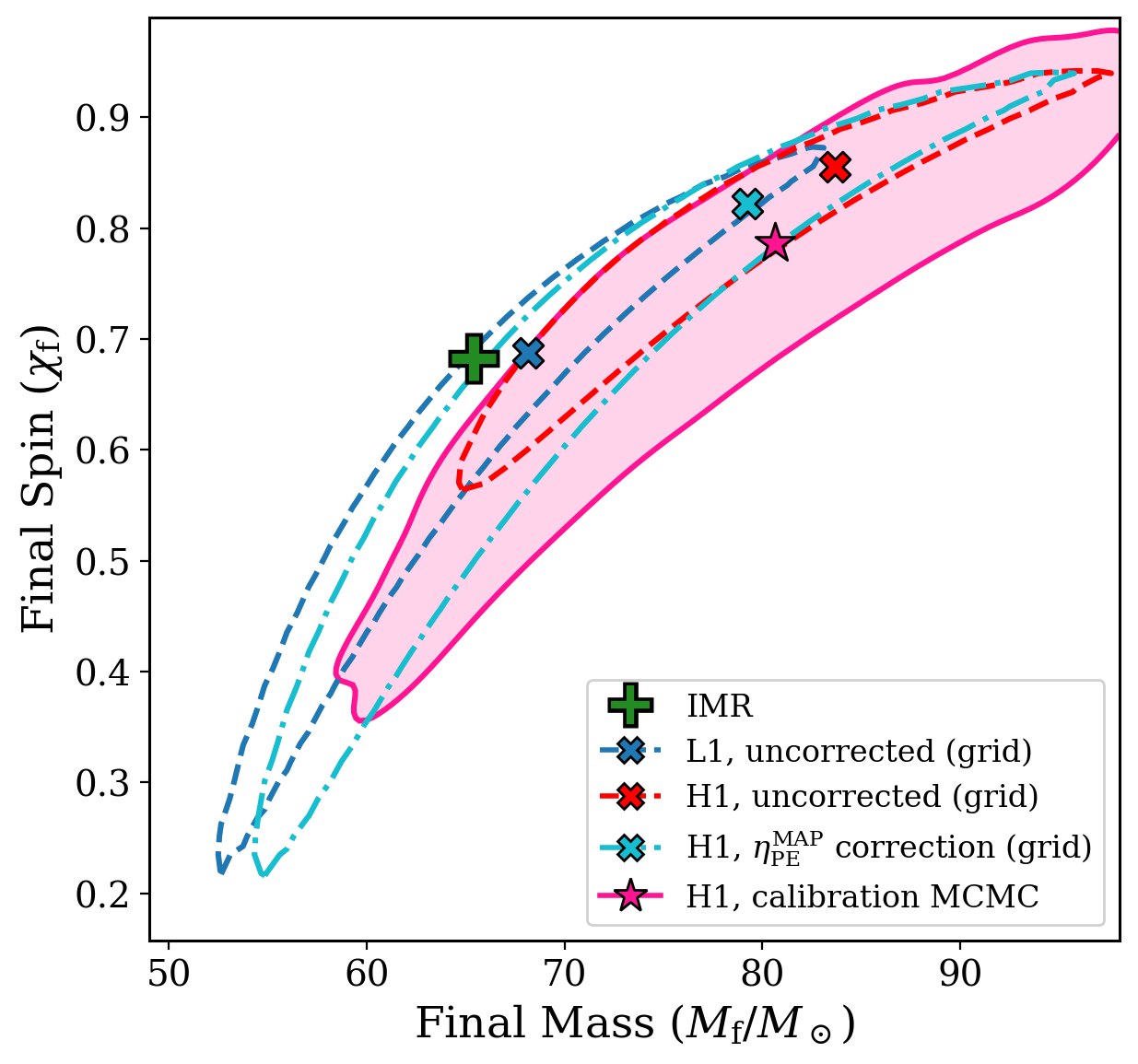}
    \caption{As \cref{fig:250207_Mchi_2det}, but for single-detector analysis using the \{220\}-only model at $t_0=8\,\MIMR$. The green plus marks the remnant predicted by the full IMR analysis. The L1-only result is shown in dark blue, while the H1-only results are shown with the calibration left uncorrected (red), corrected using $\eta_{\rm PE}^{\rm MAP}$ from \cite{astro_calib_paper} (cyan), and marginalized within our ringdown analysis (pink). Crosses denote the maximum-likelihood estimates from the Grid analyses, while the star marks the MAP estimate from the MCMC analysis. Dashed, dot-dashed, and solid contours enclose the corresponding $90\%$ credible regions.}
    \label{fig:250207_Mchi_1det}
\end{figure}

We show the single-detector mass and spin recovery at $t_0=8\,\MIMR$ with the $\{220\}$-only model in \cref{fig:250207_Mchi_1det}. The model and start time are chosen to facilitate comparison with Ref.~\cite{astro_calib_paper}. The calibration control points are evenly spaced between $\sim209.6\,$Hz and $\sim309.6\,$Hz.
As also shown in \cref{fig:250207_1det}, no calibration treatment shows overall good agreement with the IMR prediction. 
Only the L1-only posterior contains the IMR value within its 90\% CI; all of the H1-only treatments, whether uncorrected, corrected using $\eta_{\rm PE}^{\rm MAP}$, or calibration marginalized, exclude the IMR value at 90\% CI. 
The point estimate (maximum likelihood or MAP) does however slightly improve agreement with IMR when the H1 calibration error is accounted for, with QNMRF MCMC MAP estimate and $\eta_{\rm PE}^{\rm MAP}$ corrected Grid estimate closer to the IMR value than the uncorrected one. Unlike the two-detector case, calibration marginalization makes little difference to $p_{\rm IMR}$ for the single-detector analysis (also see \cref{tab:250207_Mchi_1det}), consistent with the low single-detector SNR and the lack of an independently calibrated detector limiting the ringdown sensitivity to the calibration error.

\begin{table}
    \centering
    \begin{tabular}{|l|l|c|c|c|}
        \hline
        Data & Calibration treatment & $M_\mathrm{f}$ & $\chi_\mathrm{f}$ & $p_{\rm IMR}$ \\
        \hline
        IMR (full signal) & -- & 65.38 & 0.683 & -- \\
        L1 & uncorrected & 68.15 & 0.688 & 0.838 \\
        H1 & uncorrected & 83.67 & 0.856 & 0.997 \\
        H1 & $\eta_{\rm PE}^{\rm MAP}$ correction & 79.23 & 0.822 & 0.972 \\
        H1 & calibration MCMC & 80.66 & 0.786 & 0.997 \\
        \hline
    \end{tabular}
        \caption{Maximum-likelihood (Grid) or MAP (MCMC) remnant estimates for GW250207 ($t_0=8\,\MIMR$, filter $\{220\}$) under each calibration treatment, together with the joint posterior quantile $p_{\rm IMR}$ of the IMR-predicted remnant.}
    \label{tab:250207_Mchi_1det}
\end{table}

\section{Conclusion}
\label{sec:conclusion}

We have presented two approaches for accounting for calibration errors in ringdown analysis, both built on the QNMRF. Where the error is well measured, the correction is applied to the strain before the standard analysis; where it is not, the calibration is sampled jointly with the remnant mass and spin and marginalized over an uncertainty prior. Which is appropriate depends on what is known about an event's calibration. We validate both on NR simulations in detector-like noise and apply them to GW250207, an event recorded while H1 was not fully calibrated. Accounting for the calibration error improves agreement with the inspiral-merger-ringdown prediction, depending on the detector combination, model and start time.

The improvement is not uniform. Both approaches substantially improve agreement with the IMR prediction in the joint H1--L1 analysis; however for H1 alone the recovered $p_{\rm IMR}$ is essentially unchanged. Below some detector SNR, there is simply not enough ringdown signal to constrain the mass and spin tightly and accounting for calibration neither helps nor hurts the mass--spin recovery. GW250207 is also our only real-event test so far, and a favorable one, with H1 miscalibrated while L1 was not. Characterizing the SNR floor more precisely, and extending this validation to more events would strengthen confidence in the methods.

Looking ahead, this calibration-aware analysis could be integrated into the QNMRF pipeline used in standard LVK ringdown analyses, so that calibration systematics are accounted for regularly rather than only in dedicated follow-up studies such as this one. This is increasingly important as detector sensitivity improves: as statistical uncertainties shrink with each observing run and with the next generation of detectors, calibration systematics will make up a growing share of the total error, and analyses that do not marginalize over them will become correspondingly more biased.

\begin{acknowledgments}
The authors acknowledge the Ngunnawal, Ngambri and Bunurong people, who are the Traditional Owners of the land upon which the Australian institutions stand, Australian National University and Monash University. 
This research is supported by the Australian Research Council Centre of Excellence for Gravitational Wave Discovery (OzGrav), Project Number CE230100016. L.S. is also supported by the Australian Research Council Discovery Early Career Researcher Award, Project Number DE240100206.
The authors acknowledge the support of JST ASPIRE JPMJAP2320. 
This material is based upon work supported by NSF's LIGO Laboratory which is a major facility fully funded by the National Science Foundation. The authors are grateful for computational resources provided by the LIGO Laboratory and supported by National Science Foundation Grants PHY-0757058 and PHY-0823459.
The authors used Anthropic's Claude to support code development; all results are verified by the authors.
This manuscript carries LIGO Document No. DCC-P2600537.

\end{acknowledgments}


\appendix

\section{\label{app:calib_sens}Sensitivity of the likelihood to the frequency-dependent calibration-error profile}

As noted in \cref{ssec:joint_analysis}, scaling the noise together with the signal makes the likelihood invariant under an overall rescaling of $\eta$. In the idealized limit, this cancellation holds independently at every frequency, so the likelihood carries very little information about the exact shape of the calibration-error profile, which is instead determined largely by the prior.

The residual sensitivity arises where this cancellation is not exact. In practice, the noise model is constructed from a finite stretch of data, and the calibration correction is applied to the signal and noise estimate on two slightly different frequency grids. Neither of these affects the recovery of $M_\mathrm{f}$ and $\chi_\mathrm{f}$, which come mainly from the frequency structure of the rational filter rather than from the detailed calibration profile. We confirm this with simulation-based calibration tests (\cref{ssec:ppplots}), which show statistical coverage for the recovered remnant mass and spin.

\section{\label{app:MCMC}Sampling details}

We explore the posterior of \cref{eq:full_likelihood} using the affine-invariant ensemble sampler \texttt{emcee} \cite{emcee}.
We run $\rm{nwalkers}=48$ walkers ($4\times n_{\rm dim}$). Each walker is initialized by drawing $(M_\mathrm{f},\chi_\mathrm{f})$ uniformly over the prior support and each calibration control point from a zero-mean Gaussian prior, with a small jitter added to all coordinates to prevent degenerate starts. Burn-in is set adaptively from a short pilot chain and discarded in full before the production run. We take $32{,}000$ production steps for the NR analyses and $50{,}000$ for GW250207, thinned by a factor of four, yielding $384{,}000$ posterior samples for each NR run and $600{,}000$ for the GW250207 runs.

\section{Corner plot}
\label{app:corner_plot}
\begin{figure*}
    \centering
	\includegraphics[width=\textwidth]{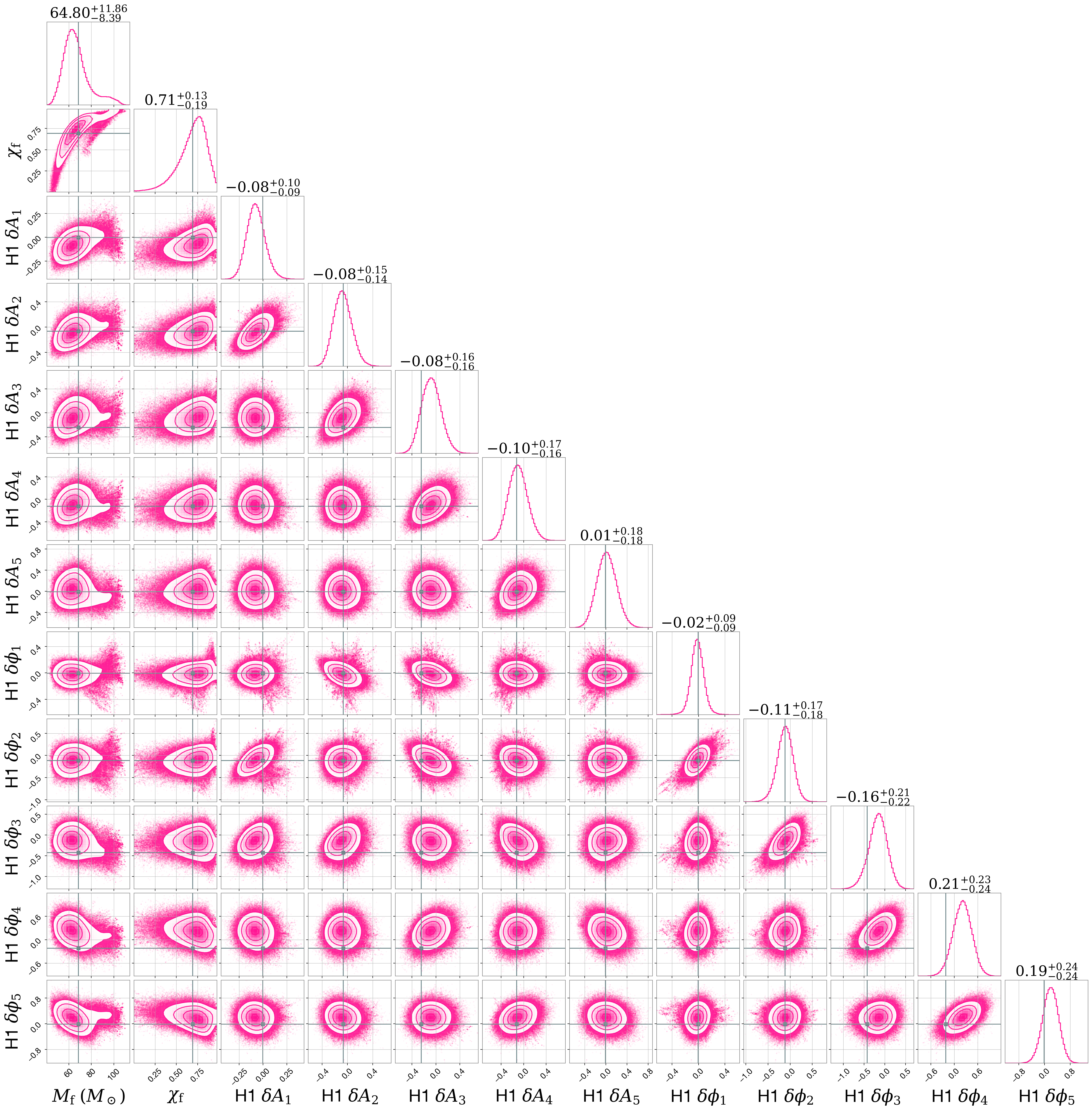}
    \caption{Corner plot of the MCMC posterior for all 12 parameters in the analysis of \cref{sssec:NR-neg}. The gray crosshairs mark the injected values.}
    \label{fig:corner}
\end{figure*}

\Cref{fig:corner} shows the full posterior for the negative-error NR analysis of \cref{sssec:NR-neg} over all 12 sampled parameters: the remnant mass and spin, and the five free amplitude and five free phase calibration control points. The recovered mass and spin are consistent with the injected values ($M_\mathrm{f}=64.80^{+11.86}_{-8.39}\,M_\odot$ compared with the injected $68.55\,M_\odot$; $\chi_\mathrm{f}=0.71^{+0.13}_{-0.19}$ compared with the injected $0.69$). The calibration control points remain broadly consistent with zero at the $1\sigma$ level, with the amplitude points showing a negative median offset that matches the sign of the simulated error. The strongest correlation is between the remnant mass and the lower-frequency amplitude control points, indicating one mechanism by which calibration uncertainty can propagate into the remnant posterior.

\bibliographystyle{apsrev4-2}
\bibliography{bib}

\end{document}